\documentclass[a4paper, fleqn, usenatbib, useAMS]{mnras}

\usepackage{graphicx}	
\usepackage{amsmath}	
\usepackage{amssymb}	
\usepackage{multicol}   
\usepackage{bm}		
\usepackage{pdflscape}	
\usepackage{subfigure}	%
\usepackage{pdfpages}
\usepackage[export]{adjustbox}
 \usepackage{ulem}
 
\usepackage{color, colortbl}
\definecolor{tableShade}{gray}{0.9}
\definecolor{red}{rgb}{0.8, 0.0, 0.0}

\newcommand{\kms}{\,km\,s$^{-1}$} 
\newcommand{\Msun}{\ensuremath{M_{\odot}}}

\newcommand{\mags}{mag~arcsec\ensuremath{^{-2}}}

\def\totalcatalogue{543~} 
\def\total{447~} 
\def\totalmemCDV{332~}
\def\memCDV{300~} 
\def\memCDVRB{272~}
\def\memCDVRB{265~}

\usepackage[T1]{fontenc}
\usepackage{ae,aecompl}

\usepackage{newtxtext, newtxmath}

\title[Antlia cluster: global properties]{
  Early-type galaxies in the Antlia Cluster: global properties
}
\author[J. P. Calder\'on et\,al.]{%
  Juan P. Calder\'on$^{1,2,3}$\thanks{E-mail:\,jpcalderon@fcaglp.unlp.edu.ar}, %
  Lilia P. Bassino$^{1,2,3}$, Sergio A. Cellone$^{1,3,4}$, %
   \newauthor
  Mat\'ias G\'omez$^{5}$ and Juan P. Caso$^{1,2,3}$\\
  $^1$Consejo Nacional de Investigaciones Cient\'ificas y T\'ecnicas, %
  Godoy Cruz 2290, C1425FQB, Ciudad Aut\'onoma de Buenos Aires, Argentina\\
  $^2$Instituto de Astrof\'isica de La Plata (CCT La Plata -- CONICET - UNLP), %
  Paseo del Bosque S/N, B1900FWA La Plata, Argentina\\
  $^3$Facultad de Ciencias Astron\'omicas y Geof\'isicas de la Universidad Nacional %
  de La Plata, Paseo del Bosque S/N, B1900FWA La Plata, Argentina\\
  $^4$Complejo Astron\'omico El Leoncito (CONICET - UNLP - UNC - UNSJ), %
  San Juan, Argentina\\
  $^5$Departamento de Ciencias F\'isicas, Facultad de Ciencias Exactas, %
  Universidad Andres Bello, Santiago, Chile\\
}
\begin{document}

\date{...}

\pagerange{\pageref{firstpage}--\pageref{lastpage}} \pubyear{2020}

\maketitle
\label{firstpage}
\begin{abstract}
We present an extension of our previous research on the early-type galaxy population of the Antlia cluster ($d \sim 35$\,Mpc), achieving a total coverage of $\sim$ 2.6\,deg$^{2}$ and performing surface photometry for $\sim$ 300 galaxies, 130 of which are new uncatalogued ones. Such new galaxies mainly fall in the low surface brightness (LSB) regime, but there are also some lenticulars (S0) which support the existence of unique functions that connect bright and dwarf galaxies in the scaling relations.
We analyse the projected spatial distribution of galaxies up to a distance of $\sim$ 800\,kpc from NGC\,3268, the adopted centre, as well as the radial velocity distribution and the correlation between galaxy colour and effective radius with the projected spatial distribution. We also obtain the luminosity function of the early-type galaxies and the distribution of stellar masses using the $T_{1}$-band magnitudes and adopted mass-luminosity ratios. Additionally, we correlate the central galaxy distribution with an X-ray emission map from the literature. 
Based on the analysis of the radial velocities and galaxy colour distributions, we find that galaxies redder than the colour-magnitude relation (CMR) have a velocity distribution strongly concentrated towards the values of the dominant galaxies and are homogeneously distributed throughout the cluster. Those bluer than the CMR, in turn, have a much more extended radial velocity distribution and are concentrated towards the centre of the cluster.
We also identify 12 candidates to ultra diffuse galaxies (UDG), that seem to be split into two families, and speculate about their origins in the context of the cluster structure.
\end{abstract}


\defcitealias{2015MNRAS.451..791C}{Paper~I}
\defcitealias{2018MNRAS.477.1760C}{Paper~II}
\defcitealias{1990AJ....100....1F}{FS90}

\section{Introduction}\label{sec:introduction}
Over the last years, low luminosity galaxies have received increasing attention from the community, due to newly discovered galaxies in nearby clusters, with unexpected surface brightness and size properties. The complexity attained by modern $\Lambda$CDM models in reference to baryonic effects \citep{2010MNRAS.406..922T, 2011MNRAS.416.1377V, 2019MNRAS.489.5742G} and the new observational evidence brought by larger detector arrays  ---MOSAIC\,II\footnote{\url{https://www.noao.edu/ctio/mosaic/}}, Dark Energy Camera \citep[DECam]{2015AJ....150..150F} and OmegaCAM Camera \citep[OmegaCAM]{2002Msngr.110...15K}--- have led the research on the distribution and properties of these newly detected galaxies. These galaxies with extreme properties represent new challenges for the theoretical models of galaxy formation and evolution.

While early-type galaxies (ETGs) in the lowest surface brightness regime have been usually classified as dwarf ellipticals (dEs), at fainter magnitudes ($M_{V} \gtrsim -16$\,mag) ETGs are represented by dwarf spheroidals \citep[dSphs,][]{2003AJ....125.1926G}, which show effective radii ($r_\mathrm{e}$) smaller than $\sim 1$\,kpc, and are supposed to have experienced intense gas and metal removal due to galactic winds \citep{2011ApJ...742L..25K}. Still lower in the luminosity scale, the ultrafaint dwarf galaxies  \citep[UFDGs,][]{2005AJ....129.2692W} are characterised for being the most dark matter dominated, the most metal-poor, and least chemically evolved stellar systems ever investigated \citep{2019ARA&A..57..375S}. The stellar systems with the lowest surface brightness, observed in nearby clusters as Virgo, Fornax, and Coma, are the so called ultra diffuse galaxies \citep[UDGs,][]{2015ApJ...798L..45V, 2018ApJ...862...82L}. They have large radii ($r_\mathrm{e} \gtrsim 1$\,kpc), moderate stellar masses, and their low surface brightnesses ($\mu_{0,V} \sim 27$\,\mags) make them only accessible to very deep observations \citep{2018ApJ...855..142E}. \cite{2016MNRAS.459L..51A} show that the origin of UDGs in clusters could be well explained by the classical model of disk formation in haloes with higher initial angular momentum.

Modern imaging surveys ---e.g.: Next Generation Virgo Cluster Survey \cite[NGVS,][]{2012ApJS..200....4F}, Fornax Deep Survey \cite[FDS,][]{2017A&A...608A.142V}, or Next Generation Fornax Survey \cite[NGFS,][]{2018ApJ...855..142E}--- allow to perform very deep multi-band photometry, covering at the same time a projected area on the sky similar to that attained by old photographic plates. This enables the study of the projected spatial distribution of the different populations that coexist in the clusters, even outside the virial radius. Despite that, at such low surface brightness levels, cluster membership through spectroscopic redshift determinations is very difficult to achieve. Anyway, such surveys have provided us with important data to study the effects of tidal interactions in galaxy evolution. The most relevant environmental effect for the evolution of early-type dwarfs is {\it harassment} \citep{1996Natur.379..613M}, which is related with repeated tidal encounters between the infalling galaxy and other cluster members or the cluster gravitational potential. This could produce a galaxy mass loss that depends on the orbit of the infalling galaxy within the cluster, and it is expected to be more efficient for low-mass galaxies \citep{2010MNRAS.405.1723S}. On the other hand, the motion of the infalling galaxy across the intracluster medium could produce the loss of the galaxy atomic gas and, consequently, the interruption of star formation \citep{1972ApJ...176....1G}.  Hydrodynamical simulations \citep[ and references therein]{2008MNRAS.388L..89R} indicate that this effect is not only due to gravitational interactions, but also to mixing turbulence and viscous effects. Additionally, an alternative effect, called {\it strangulation}, has been proposed, in which the infall of cool gas into the galaxy is stopped \citep{2000ApJ...540..113B}. This can happen when a subhalo of dark matter is created from a previously existing larger halo. The infall of gas associated to the subhalo can be cut, which results in a gradual decrease in the stellar formation, causing the galaxy colour to become redder \citep{2009ApJ...697..247W}. The result of the strangulation effect is a galaxy with higher metallicity and stellar mass than its progenitor \citep{2015Natur.521..192P}. The efficiency of those physical processes is directly related with the parameters that define the structure and dynamics of each galaxy, such as total mass, relative orbits and velocities; and also their spatial distribution inside the cluster potential. The so-called {\it morphology-density relation} \citep{1997ApJ...490..577D,   2010ApJ...714.1779V}, which locates the ETGs mainly in the central areas of the clusters, and the late-type galaxies (LTGs) in the surrounding areas, would be a result partially due to the role played by environmental processes in the evolution of galaxies. In this sense, \cite{2009ApJ...697..247W} show that both  early-type dwarf satellites and central galaxies could be affected by environmental effects in very similar ways, regardless of their locations within the cluster potential. This would indicate that the environment modifies the colour of the galaxy to a greater extent than its morphology.

The first catalogue of the Antlia Cluster was provided by \citet[][hereafter FS90]{1990AJ....100....1F} and consisted of 375 visually identified galaxies with a limiting magnitude of $B = 18$\,mag covering an area of 6.1 square degrees. Due to the lack of spectroscopy, they also provided a membership status, mainly based on morphological criteria \citep{1985AJ.....90.1759S}, assigning status 1, 2, or 3 to `definite', `likely', or `possible' members, respectively.  The two brightest ellipticals (Es) in Antlia (NGC\,3268 and NGC\,3258), located in the central region, are considered as the dominant galaxies of the cluster. This cluster exhibits a high fraction of ETGs (considering those with membership status 1 and 2): $({\rm  Es+dEs+S0s+dS0s})=(0.05+0.59+0.07+0.05)=0.76$, comparable with the Virgo and Fornax galaxy clusters, though Antlia has a central galaxy density a factor $1.4-1.7$ higher than them \citepalias{1990AJ....100....1F}. By means of X-ray observations, it has been shown that Antlia is the nearest isothermal non-cool core cluster ($kT \sim 2$\,keV, \citealp{2016ApJ...829...49W}), without a central brightness excess \citep{2000PASJ...52..623N}. NGC\,3268 is the centre of the X-ray emission, which is elongated toward a subgroup centred on NGC\,3258 \citep{1997ApJ...485L..17P, 2000PASJ...52..623N, 2008MNRAS.386.2311S}. In addition, the galaxy projected distribution presents a similar orientation, mainly in the direction defined by the two dominant Es. The Antlia cluster seems to be dynamically younger than Virgo and Fornax, but unexpectedly relaxed when it is compared to other groups \citep{2016ApJ...829...49W}, which is also suggested by optical surveys \citep[\citetalias{1990AJ....100....1F};][]{2015MNRAS.452.1617H}. The globular cluster systems of the two brightest Es have been studied by \citet[ and references therein]{2017MNRAS.470.3227C}.

In the present paper, we perform a new surface photometry of the Antlia ETGs that complements previous papers \citet[hereafter Paper I]{2015MNRAS.451..791C}, and \citet[hereafter Paper II]{2018MNRAS.477.1760C}.  Our study covers the largest area considered by any survey of this cluster until now, which motivates us to analyse the projected spatial density of galaxies up to $\sim700$\,kpc from the cluster centre, assuming an Antlia distance of $\sim35.2$\,Mpc \citep{2003A&A...408..929D}. We also recalculate the relations between structural parameters of the ETGs, now including more Es at the bright end than previous works. Moreover, we study the correlation between colour, effective radius, and radial velocity with the projected spatial distribution. Finally, we build the luminosity function of the cluster and estimate stellar masses.

This paper is organized as follows: in Sec.\,\ref{sec:data} we describe the data reduction, fit of the surface brightness profiles, membership classification, and the catalogue presented in this paper. In Sec.\,\ref{sec:results} we report the results on the new colour-magnitude relation (CMR), structural parameters, spatial projected distribution, and galaxy colour correlations. The discussion of the results is given in  Sec.\,\ref{sec:discussion} and finally, the main conclusions are summarized in Sec.\,\ref{sec:conclusions}.

\section{Observations and data reduction}\label{sec:data}
We perform a photometric study on four new fields that complements those carried out in previous surveys on the Antlia cluster of galaxies. These fields were observed with the MOSAIC\,II camera mounted on the 4-m Blanco telescope at the Cerro Tololo Inter-American Observatory (CTIO, Chile), during the nights of February 11 and 12, 2010. They correspond to the ones labelled as 4 to 7 in Table~\ref{tab:fields-coordinates}, while the fields 0 to 3 are included in the same Table for completeness, as they were studied in \citetalias{2015MNRAS.451..791C} and obtained with the same instrumental setup. As in our previous studies, images were acquired selecting the Washington $C$ and Kron-Cousins $R$ filters; observations were later transformed into the genuine Washington $C$ and $T_{1}$ bands (taking into account the zero-point offset: $R - T_{1} = -0.02$). The coverage of MOSAIC\,II images is about 8000\,$\times$\,8000\,pixels with a scale of 0.27\,arcsec/pixel, that corresponds to 36\,arcmin $\times$ 36\,arcmin or, equivalently, to 370$ \times $370\,kpc$^{2}$, according to the adopted Antlia distance. Each field corresponds to a single MOSAIC\,II image, obtained by combining between 5 and 7 dithered frames. Table~\ref{tab:fields-coordinates} presents the basic information of each field including the date of observation, position of the field centre, filter, number of exposures ($n_{f}$) that were averaged to obtain the final image for each field, individual exposure time, mean airmass ($X_{R}$ and $X_{C}$), and the seeing on the final $R$-band image. The image reduction was performed using the \textsc{MSCRED} package from \textsc{IRAF}, with the standard procedure already explained in the previous papers \citepalias[e.g.][]{2018MNRAS.477.1760C}.
\begin{table*}
\begin{minipage}{110mm}
  \caption{Observational data for the MOSAIC\,II fields (see text).}
    \label{tab:fields-coordinates}
    \begin{tabular}{@{}ccccccccc}
    \hline
    Field & Observation & $\alpha_{\rm 2000}$ & $\delta_{\rm 2000}$ &
    Filter & $n_{\rm f}$ & Exposure & Airmass & FWHM \\
           & date        &          &         &
           &             &  [s]     &         & [arcsec] \\
\hline
0 & April 2002 & 10:29:22 & $-$35:27:54 & $R$ & 5 &  600 & 1.059 & 1.0 \\ \rowcolor{tableShade}
  &            &          &             & $C$ & 7 &  600 & 1.037 & 1.1 \\ 
1 & March 2004 & 10:28:59 & $-$34:57:40 & $R$ & 5 &  600 & 1.588 & 1.0 \\ \rowcolor{tableShade}
  &            &          &             & $C$ & 7 & 1000 & 1.076 & 1.0 \\ 
2 & March 2004 & 10:31:09 & $-$34:55:59 & $R$ & 5 &  600 & 1.056 & 1.0 \\ \rowcolor{tableShade}
  &            &          &             & $C$ & 7 &  900 & 1.016 & 1.2 \\ 
3 & March 2004 & 10:31:35 & $-$35:30:42 & $R$ & 5 &  600 & 1.269 & 0.9 \\ \rowcolor{tableShade}
  &            &          &             & $C$ & 7 &  900 & 1.030 & 0.8 \\ 
  \hline
4 & February 2010 & 10:27:01 & $-$35:30:42 & $R$ & 5 &  600 & 1.106 & 1.5 \\ \rowcolor{tableShade}
  &               &          &             & $C$ & 6 &  900 & 1.025 & 1.9 \\ 
5 & February 2010 & 10:31:39 & $-$36:06:59 & $R$ & 5 &  600 & 1.278 & 1.1 \\ \rowcolor{tableShade}
  &               &          &             & $C$ & 6 &  900 & 1.023 & 1.6 \\ 
6 & February 2010 & 10:29:38 & $-$35:56:19 & $R$ & 5 &  720 & 1.259 & 1.5 \\ \rowcolor{tableShade}
  &               &          &             & $C$ & 7 &  900 & 1.080 & 1.6 \\ 
7 & February 2010 & 10:26:58 & $-$35:55:22 & $R$ & 5 &  600 & 1.025 & 1.2 \\ \rowcolor{tableShade}
  &               &          &             & $C$ & 6 &  900 & 1.503 & 1.6 \\
\hline
\end{tabular}
\end{minipage}
\end{table*}
Regarding the calibration to the standard system, we obtained the following transformation equations for the new fields (4, 5, 6, and 7) to transform instrumental magnitudes ($m_{R}$ and $m_{C}$) to standard ones ($T_{1}$ and $C$), based on standard stars fields from \cite{1996AJ....111..480G}:
\begin{eqnarray}\label{eq:calibration_equation}
  T_{1} & = & (m_{R} + 0.02 ) + a_{1} + a_{2}\,X_{R} +  a_{3}\,(C-T_{1}) \\
  C & = & m_{C} + b_{1} + b_{2}\,X_{C} +  b_{3}\,(C-T_{1}) \textnormal{,}\nonumber
\end{eqnarray}
where the coefficients and their errors are given in Table\,\ref{tab:calibration_coefficient}. For the rest of the fields, we used the corresponding calibration equations given by \citet{2003A&A...408..929D} for the central field, and \citetalias{2015MNRAS.451..791C} for fields 1, 2, and 3. In addition, we applied a zero-point magnitude offset to refer the photometry to the central field, as explained in \citetalias{2015MNRAS.451..791C}. The offsets in the $T_{1}$-band are between $\sim 0.06$\,mag and $\sim 0.1$\,mag, while in the $C$-band they are $\sim 0.2$\,mag
\begin{table}
  \caption{Coefficients (top row) and errors (lower row) for the calibration  equations\,\ref{eq:calibration_equation}.}\label{tab:calibration_coefficient}
  \centering
  \begin{tabular}{cccccc}
    \hline
    $a_{1}$  & $a_{2}$  & $a_{3}$   & $b_{1}$ & $b_{2}$ & $b_{3}$ \\
    \hline
    0.466 & $-0.063$ & 0.031 & 0.029 & $-0.405$ & 0.096 \\ \rowcolor{tableShade}
    0.026 & $\phantom{-}0.018$ & 0.007  & 0.059 & $\phantom{-}0.052$ & 0.007 \\
    \hline
  \end{tabular}
\end{table}
All magnitudes and colours depicted in the Figures shown in this paper have been corrected for Galactic absorption and reddening. We used the colour excess $E(B -V)$ from \cite{2011ApJ...737..103S}, and the relations $E(C-T_{1}) = 1.97 E(B-V)$ \citep{1977AJ.....82..798H} and $A_{R}/A_{V} = 0.75$ \citep{1985ApJ...288..618R} in order to apply the corrections corresponding to the Washington photometric system.

\subsection{Photometry of the Antlia Cluster}\label{sec:from_plate}
\begin{figure*}
  \centering
  \includegraphics[width=0.95\textwidth]{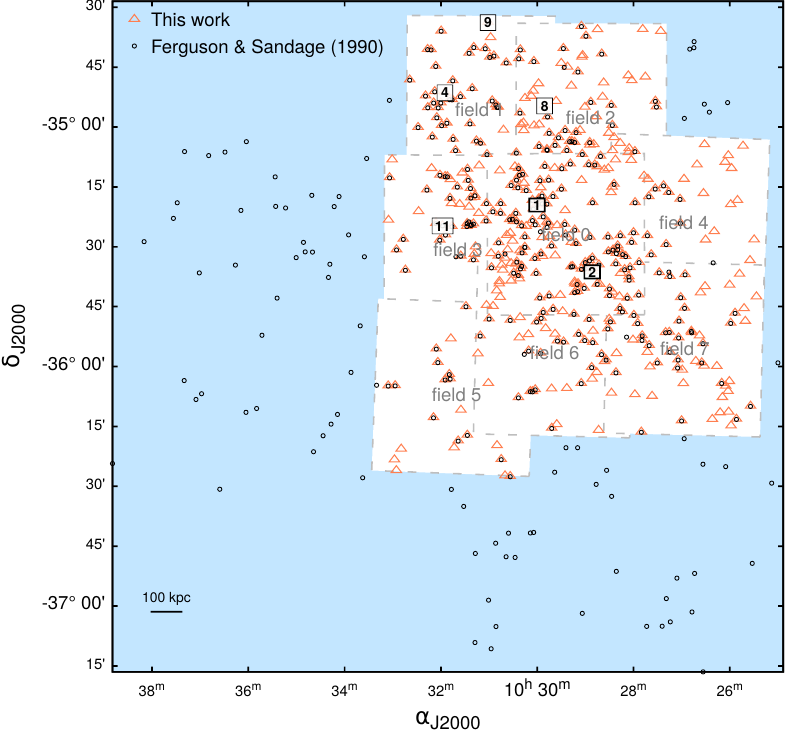}
  \caption{Position of the galaxies from the original FS90 catalogue (open circles) and those studied in the present paper (from FS90 and newly discovered ones, open triangles). Numbers 1 and 2 (within field 0) identify the two brightest elliptical galaxies NGC\,3268 and NGC\,3258, respectively, and numbers 4, 8, 9, and 11 (outside field 0) identify other galaxies brighter than $T_{1} = 20$ (see Table\,\ref{tab:topmagnitude} for their identifications and main properties).}\label{fig:radec-fs90}
\end{figure*}

The first paper presenting a CCD study of the Antlia cluster by \citet{2008MNRAS.386.2311S}, was based on the original Antlia catalogue \citepalias{1990AJ....100....1F}, which involved 375 galaxies from early to late-type morphologies. In Fig.\,\ref{fig:radec-fs90}, we compare the spatial distribution of the galaxies included in the FS90 catalogue (open circles) with the sample studied in the present work (open triangles). We labelled each MOSAIC\,II field from `field 0' (the central one) to `field 7'. Their central coordinates are also given in Table\,\ref{tab:fields-coordinates}.

The central field (0) was used in \cite{2008MNRAS.386.2311S} to study 100 ETGs from the FS90 catalogue through isophotal photometry, while the existence of two compact elliptical (cE) galaxies in that sample was confirmed by \cite{2008MNRAS.391..685S}. Then, new dE candidates (not previously catalogued) and radial velocities to confirm membership were added in \cite{2012MNRAS.419.2472S}. In \citetalias{2015MNRAS.451..791C} and \citetalias{2018MNRAS.477.1760C}, we extended the galaxy sample by adding fields 1 to 3, calculating new total magnitudes and colours ---against just isophotal photometry--- of 107 additional cluster members plus 31 candidates with high membership probability. Their structural parameters, obtained by the fits of S\'ersic models \citep{1968adga.book.....S}, were also studied, and the corresponding scaling relations derived and compared with nearby clusters. 
In the present work, we add the surface photometry of galaxies in fields 4 to 7, so that our final catalogue contains \total galaxies (\totalmemCDV of them considered as members or candidates), which constitutes the largest sample of galaxies in the Antlia cluster with homogeneous integrated photometry.

The surface photometry analysis that we used has been extensively explained in \citetalias{2015MNRAS.451..791C} and \citetalias{2018MNRAS.477.1760C}. Here we just include the main formulation applied to obtain magnitudes and structural parameters. We obtained the ETG surface brightness profiles through \textsc{ellipse} task within \textsc{IRAF}, then fitting each of them with a one component S\'ersic model:

\begin{equation}\label{eq:modelo_de_sersic_magnitudes}
  \mu(r) = \mu_\mathrm{0} + 1.0857
  \left(\frac{r}{r_\mathrm{0}}\right)^{1/n}\textnormal{,}
\end{equation}
where $\mu_\mathrm{0}$ is the central surface brightness, $r_\mathrm{0}$ is a scale parameter and $n$ is the S\'ersic index which is a measure of the concentration of the profile. From these parameters, we obtained the effective radius ($r_\mathrm{e}$), effective surface brightness ($\mu_\mathrm{e}$) and total magnitudes and colours of the galaxy sample. Each profile was fitted using the task \textsc{nfit1d} from \textsc{IRAF}, which implements the $\chi^{2}$ statistic test by the Levenberg-Marquardt algorithm. In most cases, the residuals between model and observed profile turned out to be  smaller than $0.5$\,mag \citepalias{2018MNRAS.477.1760C}. Finally, as the parameters obtained from the fits were not affected by seeing for models with $n \lesssim 3$ (which is the general case), we did not apply further corrections \citepalias{2018MNRAS.477.1760C}. 

With the aim of making a complete catalogue of galaxies present in our images, we also included the surface photometry of all late-type galaxies found on them. In these cases, we estimate a total magnitude, but the one-component scale parameters derived may differ from those of a more complex fit, with different components involved. In any case, late-type galaxies were not included in the fits, regressions or statistical tests (unless explicitly indicated). 

\subsection{Membership status of the ETGs}
\begin{figure}
  \centering
  \includegraphics[width=0.95\columnwidth]{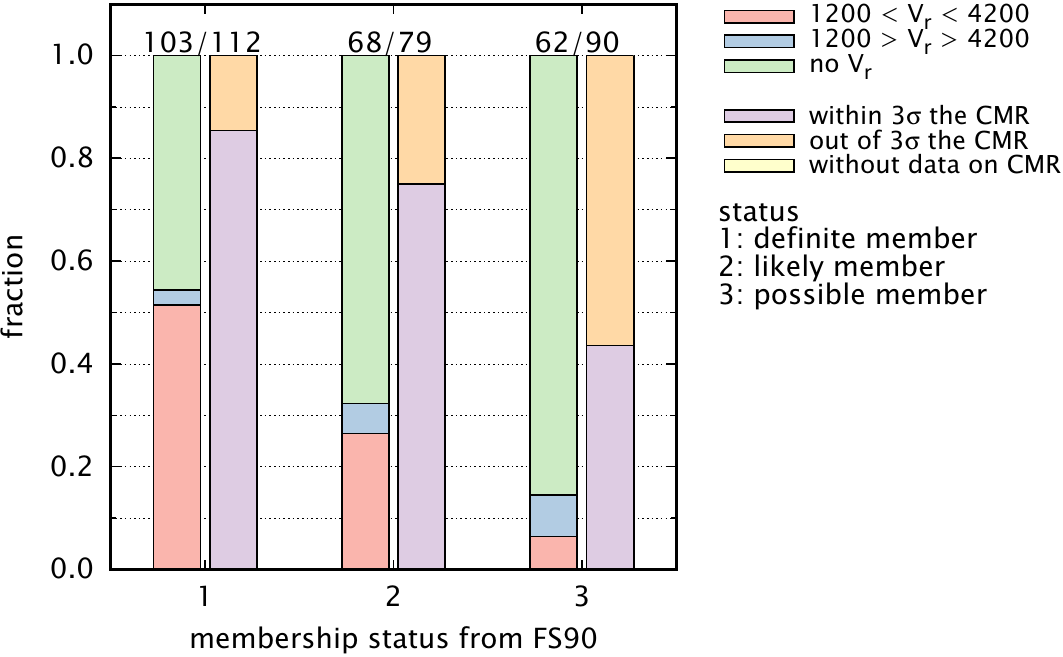}
  \caption{Comparison between the membership classification of ETGs: FS90 status, radial velocities, and `$\pm 3 \sigma$ from the CMR'  criterion. For each membership status, the right column corresponds to the CMR criterion and the left column to the radial velocity measurements.}\label{fig:h-mem_vr}
\end{figure}

The membership status applied by FS90 was mainly based on morphological criteria due to the lack of spectroscopic data. Afterwards, it was proved with radial velocities that the FS90 status is in fact highly reliable \citep{2012MNRAS.419.2472S, 2015MNRAS.451..791C}. In our series of papers on the Antlia cluster, the radial velocity range we have considered to define confirmed members is $1200 - 4200$\,\kms \citep{2008MNRAS.386.2311S}. We have radial velocity measurements for 50 per\,cent of the ETGs with membership status 1 and 27 per\,cent status 2; out of them, the 95 and 82 per\,cent are confirmed members, respectively. In Fig.\,\ref{fig:h-mem_vr} we show an histogram to compare the three different membership criteria that were applied to  ETGs: {\it i.} status defined by \citetalias{1990AJ....100....1F}, {\it ii.} measured radial velocities \citep{2015A&A...584A.125C}, and {\it iii.} position in the colour-magnitude diagram, i.e. considering as members those galaxies located within $\pm 3\sigma$ of the mean CMR defined by ETGs (see Fig.\,\ref{fig:CT1_T1} and  Sec.\,\ref{sec:color_magnitud}), being this method extensively used to identify cluster members \citep[e.g.][]{2012A&A...538A..69L}. 
From now on, we will refer to this latter criterion as `3SmC': $\pm 3\sigma$ mean colour criterion.

The original FS90 catalogue has 112 ETGs with membership status 1, 79 with status 2, and 90 with status 3. Out of this total of 281 ETGs: 9 of status 1, 11 of status 2, and 28 of status 3 fall outside our MOSAIC\,II fields. Additionally, the identifications FS90 203/206 represent the same galaxy, as well as FS90 269/270. Several FS90 galaxies could not be found on the MOSAIC\,II images: FS90 11, 42, 74, 151; and there are five galaxies with doubtful coordinates (big shifts from FS90): FS90 178, 275, 204, 235, 287. When analysing just the ETGs of the FS90 sample on the MOSAIC\,II images, we found that out of the 103 galaxies catalogued as `definite' members, more than 80\,per cent follow the 3SmC criterion, while almost $\sim 50$\,per cent of them have radial velocities that confirm them as cluster members. Comparatively, galaxies in the two remaining membership statuses maintain a similar relation between those being confirmed by radial velocities and by the 3SmC criterion. Considering the 68 `likely' members (status 2), more than 65\,per cent are members following the 3SmC criterion, while 25\,per cent have spectroscopic confirmation. Lastly, out of the 62 `possible' members, 40\,per cent follow the 3SmC criterion, and only about 5\,per cent are spectroscopic confirmed members. These results (see also Fig.\,\ref{fig:h-mem_vr}) reinforce the validity of the FS90 membership status. 

According to the previous results, we can consider ETGs with FS90 membership status 1 plus those confirmed spectroscopically as fully confirmed members. In order to analyse quantitatively if the ETGs with membership status 2 can also be included in that sample of confirmed members \citep{2012MNRAS.420.3427B, 2017A&A...608A.142V}, we calculated  their residuals as the difference in magnitude and colour between their positions in the colour-magnitude diagram with respect to the mean CMR. Then, we performed a Kolmogorov-Smirnov (KS) test \citep{1992nrca.book.....P} to compare the distribution of such residuals between two samples: the 17 ETGs with FS90 status 2 that are confirmed as members through radial velocities on one side, and the whole sample of 68 ETGs with FS90 status 2 on the other. As a result of the test we obtained a KS statistic of 0.25 and a probability $1-p = 0.68$. Thus, we cannot reject the hypothesis that the statistical distributions of the residuals of both samples, in the colour-magnitude diagram, come from a common distribution. On the basis of these results, we decided to add the ETGs with FS90 membership status 2 to the sample of `confirmed members'.

Additionally, we calculated the colour residuals with respect to the CMR for all ETGs in the confirmed member sample \citep[e.g.][]{2019A&A...625A..94H}, and fitted a Gaussian distribution to the resulting histogram (not shown). Then, we filtered out seven galaxies lying further than $3\sigma_{CT_1}$ of the Gaussian mean (with $\sigma_{CT_1} = 0.17$). Although these galaxies are confirmed members, their colour profiles seem to have unexpectedly large errors, so we decided not to take them into account for any further analysis.

\subsection{The catalogue}\label{sec:the_catalog}
The total number of galaxies identified in all fields is \total (including FS90 and newly discovered galaxies of all morphologies). Out of them, we have obtained surface photometry for \memCDV (130 are new ones, uncatalogued in previous papers) which meet any of the following requirements: as ETG and LTG `confirmed members' they have been confirmed by radial velocity or assigned FS90 membership status 1 or 2, or as just ETG `candidates' they follow the 3SmC criterion (see \citetalias{2015MNRAS.451..791C}). In Table\,\ref{tab:new_antlia} we present their photometric and structural parameters, with an asterisk indicating those galaxies that have been studied for the first time in the present paper.

Finally, there are \memCDVRB ETGs included in the catalogue, that we will call the {\it final sample}, i.e. ETG confirmed by radial velocity or assigned FS90 membership status 1 or 2, plus ETG `candidates' that follow the 3SmC criterion. 

\begin{table*}
\caption{Catalogue of galaxies in the Antlia  cluster. The columns are described as follows: (1) id from FS90 or new IAU identification, (2)-(3) J2000 coordinates, (4) morphology from FS90, (5) membership status (from FS90 except those with radial velocity), (6) Galactic extinction from \citet{2011ApJ...737..103S}, (7)-(12) global properties calculated in this work: $T_{1}$-band magnitude, $(C - T_{1})$ colour, $T_{1}$-band effective surface brightness, effective radius, S\'ersic index, stellar mass (see Sec\,\ref{sec:stellar_masses}), (13) radial velocity (from \citealt{2012MNRAS.419.2472S, 2015A&A...584A.125C, 2009MNRAS.399..683J} or NED\protect\footnotemark).}\label{tab:new_antlia}
\setlength{\tabcolsep}{0.6ex}
\begin{tabular}{cccccccccccccl}
  id & $\alpha_{\mathrm{J}2000}$ & $\delta_{\mathrm{J}2000}$ & morph. & mem & $E(B-V)$ & $T_{1}$ & $(C-T_{1})$ & $\mu_\mathrm{e}$ & $r_\mathrm{e}$ & $n$ & $\log(M_{\mathcal{G}})$ & $V_{R}$ & Notes\\
     &                &                 &        &     & [mag]    & [mag]  & [mag]       & [\mags]  & [kpc]  &     &  [\Msun]             & [\kms]   &   \\
\hline
 {\scriptsize 1}                      & 10:25:05.040 & -35:58:58.800 & SmV               & 1 & ----- & ----- & ---- & ----- & ----- & ----- & ----- &  3220$\pm$05 & FS-\,-\, \\  \rowcolor{tableShade}
 {\scriptsize 2}                      & 10:25:10.800 & -36:29:09.600 & S or ImV          & 3 & ----- & ----- & ---- & ----- & ----- & ----- & ----- & ------- & F-\,-\,-\, \\
 {\scriptsize ANTLJ102532.4-354233.7} & 10:25:32.424 & -35:42:33.721 & S0                & -- & 0.065 & 15.55 & 2.04 & 21.98 &  0.92 &  1.17 & 11.19 & ------- & -\,-\,-\,C* \\  \rowcolor{tableShade}
 {\scriptsize 3}                      & 10:25:33.600 & -36:49:15.600 & dE                & 3 & ----- & ----- & ---- & ----- & ----- & ----- & ----- & ------- & F-\,-\,-\, \\
 {\scriptsize ANTLJ102536.1-352816.3} & 10:25:36.115 & -35:28:16.311 & dE                & -- & 0.066 & 18.83 & 2.45 & 22.61 &  0.27 &  1.14 &  9.95 & ------- & -\,-\,-\,C* \\  \rowcolor{tableShade}
 {\scriptsize 4}                      & 10:25:38.147 & -36:09:56.289 & ImV or S          & 3 & 0.068 & 16.49 & 1.19 & 22.53 &  0.79 &  1.08 & 10.46 & ------- & F-\,-\,C* \\
 {\scriptsize ANTLJ102540.7-353326.6} & 10:25:40.718 & -35:33:26.610 & dE                & -- & 0.062 & 18.57 & 2.36 & 22.72 &  0.37 &  0.60 & 10.05 & ------- & -\,-\,-\,C* \\  \rowcolor{tableShade}
\hline
\end{tabular}
\raggedright Notes.- The last column gives the references of the papers where the galaxy has been studied. The code FSCC refers to: (F) \cite{1990AJ....100....1F}; (S) \cite{2008MNRAS.386.2311S} and \cite{2012MNRAS.419.2472S}; (C) \cite{2015A&A...584A.125C}; (C) \cite{2015MNRAS.451..791C} and \cite{2018MNRAS.477.1760C}. Asterisks identify galaxies studied for the first time in the present paper. The full table can be accessed electronically.
\end{table*}
\footnotetext{The NASA/IPAC Extragalactic Database (NED) is operated by the Jet Propulsion Laboratory, California Institute of Technology, under contract with the National Aeronautics and Space Administration.}

\section{Results}\label{sec:results}
\subsection{Colour-magnitude relation}\label{sec:color_magnitud}
\begin{figure}
  \begin{center}
      \includegraphics[width=0.45\textwidth]{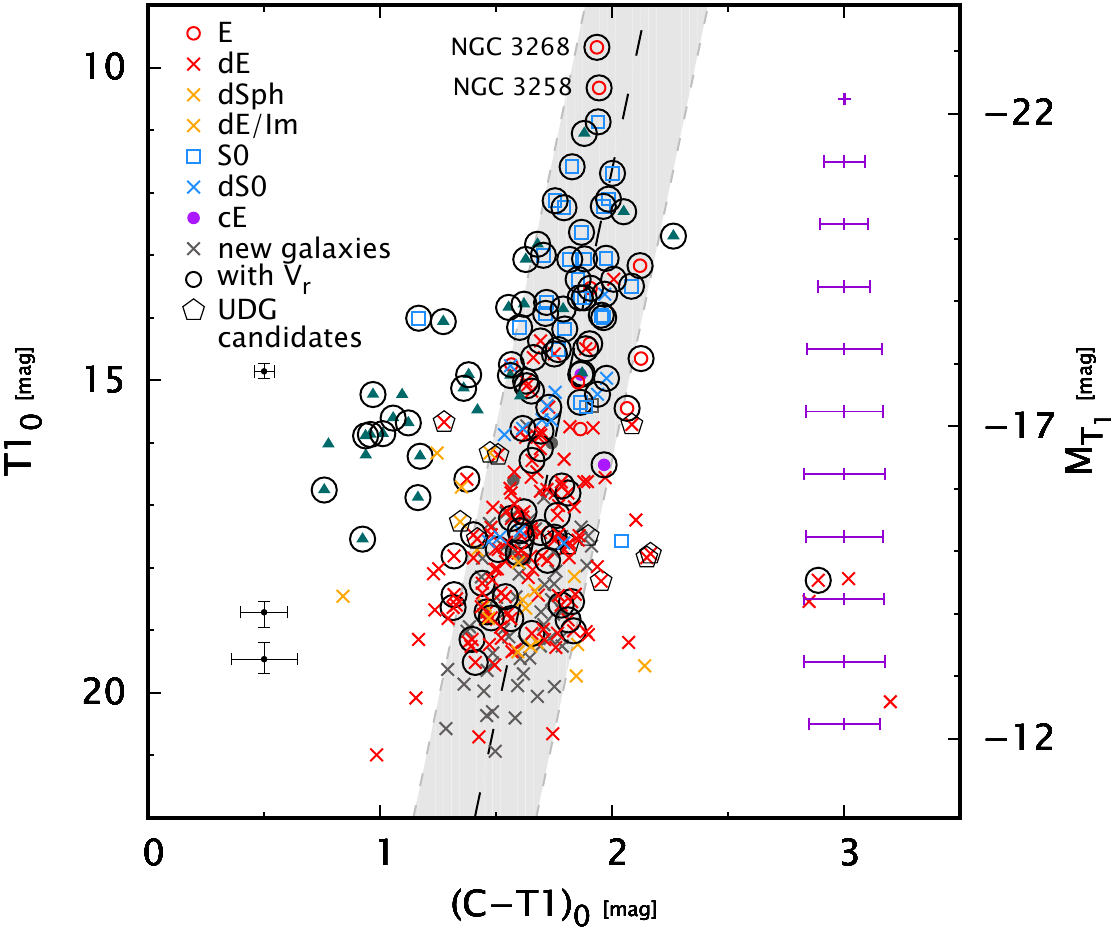}%
      \caption{Colour-magnitude diagram for the ETGs in the final sample (i.e. confirmed members plus candidates following the 3SmC criterion). The grey band  represents $\pm 3 \sigma$ from the fitted CMR and colour codes for the symbols are given in the plot. On the left, we show errors in colour and magnitude. On the right, the colour dispersion of the ETG final sample, for each magnitude bin. Additionally, we include LTGs identified with green triangles.}\label{fig:CT1_T1}
  \end{center}
\end{figure}

Fig.\,\ref{fig:CT1_T1} shows the $(T_{1})_{0}$ vs. $(C-T_{1})_{0}$ colour-magnitude diagram of the ETG final sample from our catalogue. We also included the LTGs represented by green triangles, that have not been taken into account for any further statistics. The black dashed line shows the mean CMR calculated through a least-square fit of the ETG `confirmed members', taking into account errors in both axes, giving:
\begin{equation}\label{eq:regression}
  (T_{1})_{0} = (-17.5 \pm 0.2)~(C-T_{1})_{0} + (46.7 \pm 3.3).
\end{equation}
The standard deviation of the fit is $\sigma=1.54$; the $\pm 3\sigma$ region about the CMR has been represented as a grey band in Fig.\,\ref{fig:CT1_T1}. Almost all ETGs confirmed as members by radial velocities, identified in Fig.\,\ref{fig:CT1_T1} with black open circles, fall within $\pm 3 \sigma$ from the mean CMR. 
The main characteristics of the CMR defined by ETGs in the Antlia cluster have been already discussed in \citetalias{2018MNRAS.477.1760C}. Now, with the addition of galaxies located in the new fields, we fill in the brighter side of the CMR and are able to see more clearly the `break' at $M_{T_1} \sim -20$\,mag \citep{2011MNRAS.417..785J}, from which brighter galaxies keep an almost constant colour. Moreover, as most of the galaxy population lies within $\pm 3 \sigma$ of the fit, the colour dispersion remains almost constant towards the faintest galaxies, as shown on the right side of Fig.\,\ref{fig:CT1_T1}. Twelve UDG candidates have been identified through their particular structural parameters (see next section for a detailed description).

\subsection{Structural parameters}
\begin{figure*}
  \begin{center}
      \subfigure[]{%
      \label{fig:Mv_mu0}
      \includegraphics[width=0.31\textwidth]{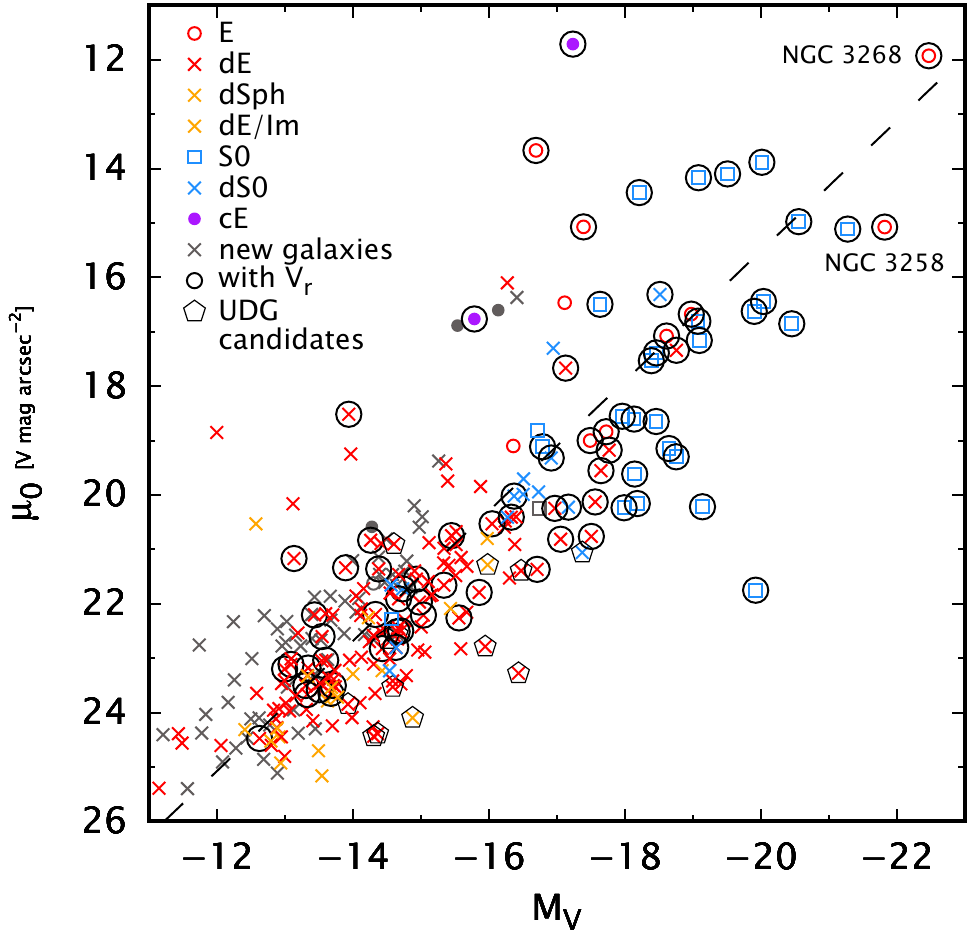}}\hspace{0.015\textwidth}
     \subfigure[]{%
      \label{fig:n_Mv}
      \includegraphics[width=0.31\textwidth]{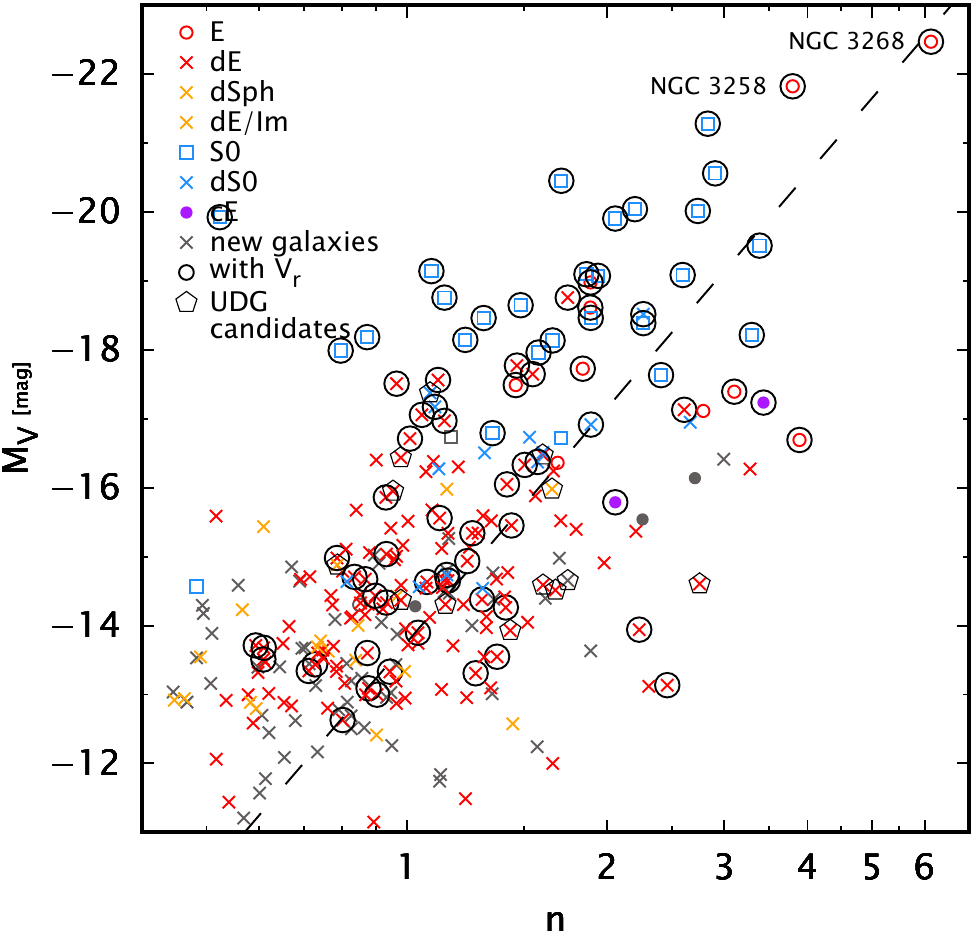}}\hspace{0.015\textwidth}
     \subfigure[]{%
      \label{fig:n_mu0}
      \includegraphics[width=0.31\textwidth]{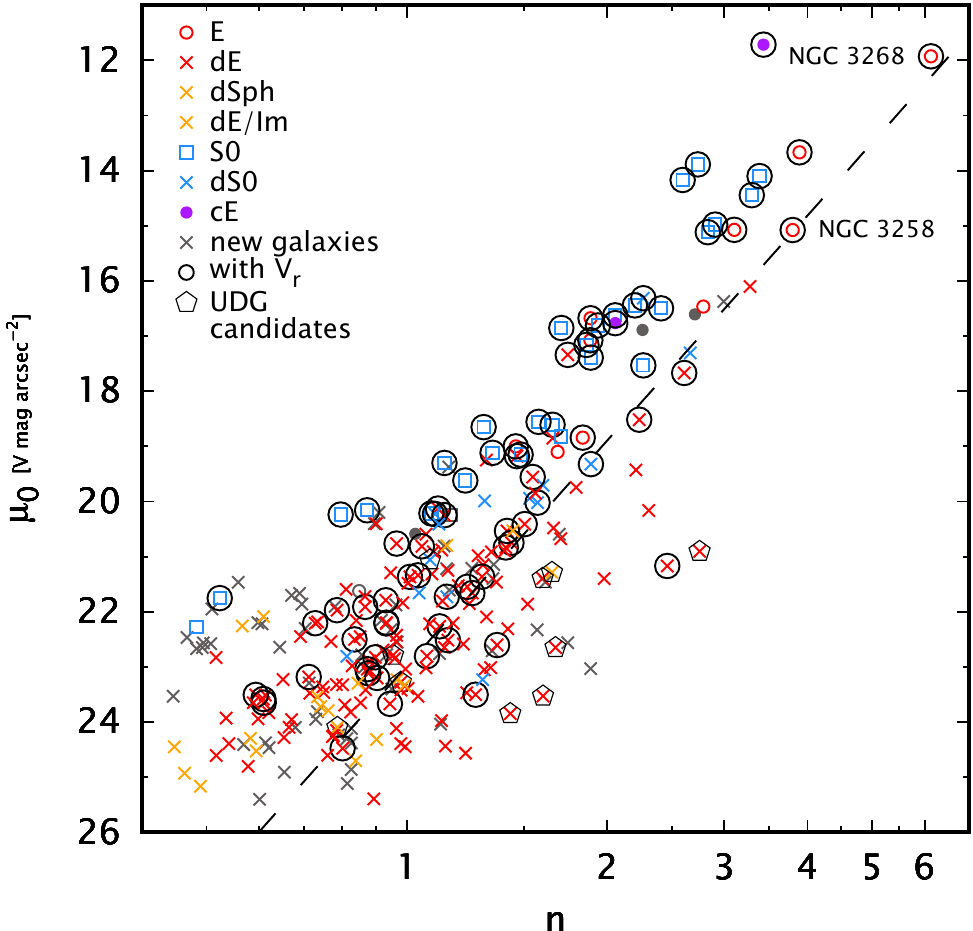}}\\
     \subfigure[]{%
      \label{fig:Mv_mue}
      \includegraphics[width=\columnwidth]{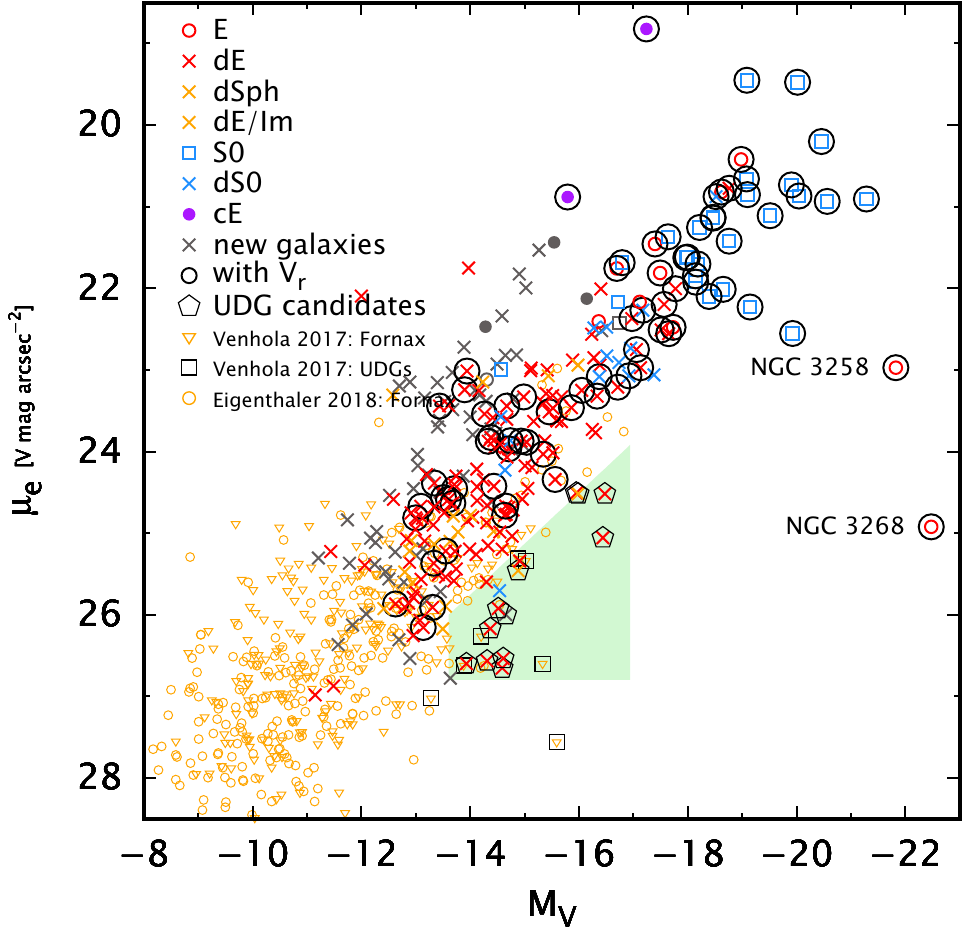}}%
     \subfigure[]{%
      \label{fig:Mv_re}
      \includegraphics[width=\columnwidth]{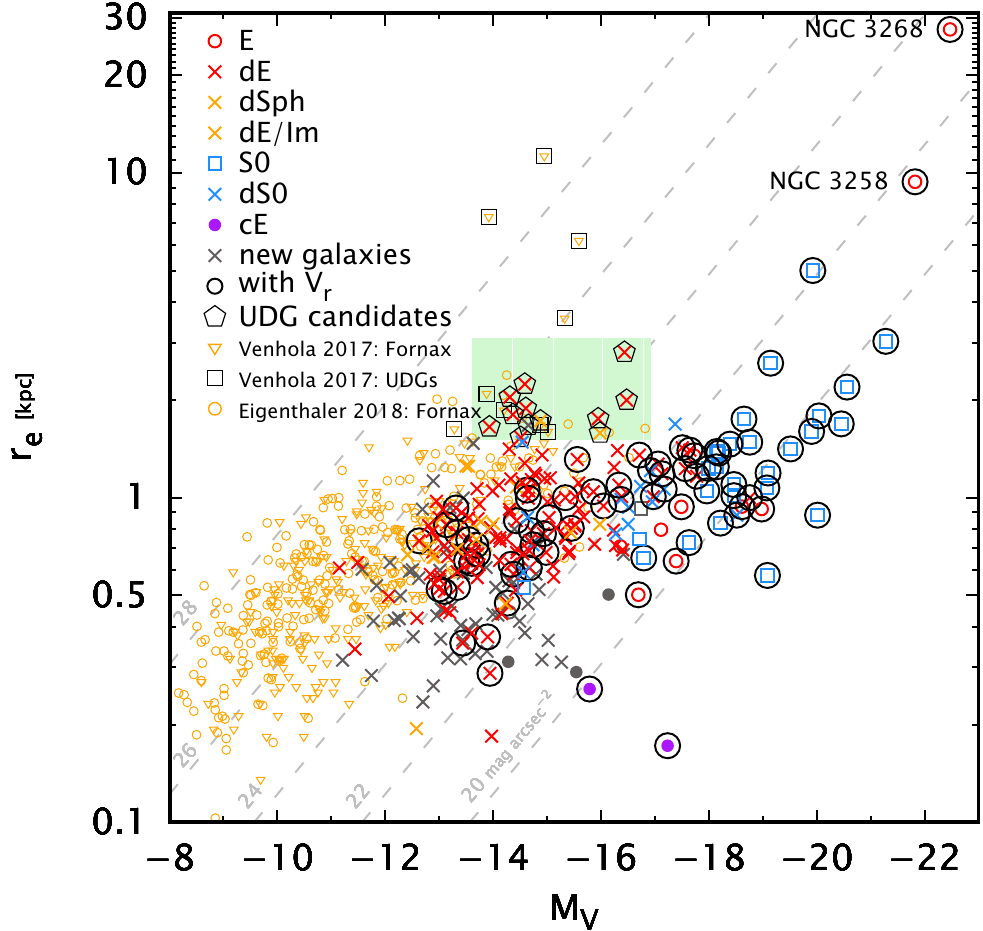}}%
     \caption{Relations between the structural parameters for the ETG final sample. Symbols are identified in each plot. Upper panels: (a) Central surface brightness versus absolute magnitude ($V$-band). (b) Absolute magnitude versus logarithm of the S\'ersic index ($V$-band). (c) Central surface brightness versus logarithm of the S\'ersic index ($V$-band). Dashed lines show the respective least-square linear fits, except in (c) that is obtained from (a) and (b) fits (see text). Lower panels: (d) Effective surface brightness versus absolute magnitude ($V$-band). (e) Logarithm of the effective radius versus absolute magnitude ($V$-band). With grey dashed lines we indicate the loci of constant mean effective surface brightness ($\langle\mu_\mathrm{e}\rangle$). Green areas in (d) and (e) highlight the position of Antlia UDG candidates.}%
     \label{fig:scaling_relations}
  \end{center}
\end{figure*}

Using the data obtained from the new MOSAIC\,II fields (4 to 7), we  have revisited the relations between structural parameters of the ETG population that involve: central surface brightness, absolute magnitude, and S\'ersic index.
Following the same procedure as in \citetalias{2015MNRAS.451..791C} and references therein, we obtained least-square fits for the linear regressions of $\mu_{0}$ vs. $M_{V}$ (Fig.\,\ref{fig:Mv_mu0}), and $M_{V}$ vs. $n$ (Fig.\,\ref{fig:n_Mv}). The linear fits were obtained taking into account numerical errors in both axes and excluding the cEs of the sample. The resulting relations are:
\begin{eqnarray}
\mu_{0} &=& (39.41 \pm 1.03) + (1.19 \pm 0.06)~M_{V} \\
M_{V} &=& (-13.75 \pm 0.88) + (-11.29 \pm 4.21)~\log(n){.}
\end{eqnarray}

By combining equations (4) and (5), we obtain the following linear relation between $\mu_{0}$ and $n$ (Fig.\,\ref{fig:n_mu0}):
\begin{eqnarray}
  \mu_{0} &=& 23.04 - 13.43~\log(n){.}
\end{eqnarray}
Based on these linear fits, we have rebuilt the relations that involve the effective parameters: effective surface brightness ($\mu_{e}$) and effective radius ($r_\mathrm{e}$), which are shown in Fig.\,\ref{fig:Mv_mue} and Fig.\,\ref{fig:Mv_re}, respectively.

In \citetalias{2015MNRAS.451..791C}, we compared the structural parameters of the Antlia galaxy population with those of other clusters and groups. The recent releases of the Next Generation Fornax Survey \cite[NGFS,][]{2018ApJ...855..142E} and Fornax Deep Survey \cite[FDS,][]{2017A&A...608A.142V}, allow us to compare our result with those from ETGs of nearby clusters.
The optical side of the NGFS is based on deep multi-band observations ($u'$, $g'$ and $i'$-bands), taken with the Dark Energy Camera (DECam,  \citealt{2015AJ....150..150F}), mounted on the 4-m Blanco telescope at the CTIO. The FDS survey, in turn, used the ESO VLT Survey Telescope (VST), a  2.6-m wide-field optical telescope at Cerro Paranal, Chile. The observations were taken in $u$, $g$, $r$, and $i$-bands with the OmegaCam camera \citep{2002Msngr.110...15K}. Despite that NGFS achieved a larger coverage than FDS, both surveys span almost the same magnitude range: $-16 < M_{V} < -8$\,mag \cite[magnitude transformations were performed according to][]{1995PASP..107..945F}, and list in their respective catalogues the total magnitude, S\'ersic index and effective radius of each galaxy. From these quantities, we calculated $\mu_\mathrm{e}$ using the following expressions:
\begin{eqnarray}
  \mu_{0} &=& m + 1.995450 + 5  \log(r_\mathrm{e}) \\
         & & - 5 n \log(b_{n}) + 2.5 \log(n~\Gamma(2 n))  \nonumber \\
  \mu_\mathrm{e} &=& \mu_{0} + 1.0857~b_{n}\textnormal{,}
\end{eqnarray}
where $\Gamma(x)$ is the complete gamma function, and $b_{n}$ can be obtained by solving $\Gamma(2n) = 2~\gamma(2n,b_n)$ \citep{1991A&A...249...99C}, where $\gamma(a,x)$ is the incomplete gamma function. In Figs.\,\ref{fig:Mv_mue} and\,\ref{fig:Mv_re} we have included data from the NGFS and FDS catalogues, with different symbols identifying the UDGs \citep{2017A&A...608A.142V}, that have larger radii than dwarf galaxies. The data from the Fornax samples extend the fainter limit of our Antlia data in about 3 magnitudes.

We have searched for UDGs in Antlia, taking into account their  characteristics as observed in Virgo, Fornax, and Coma clusters \citep{2015ApJ...798L..45V, 2017A&A...608A.142V}. If we consider as UDG candidates all dwarf galaxies with $r_\mathrm{e} > 1.5$\,kpc, then the UDGs in Antlia attain a total number of 12 candidates. These galaxies have a mean effective radius $\langle r_\mathrm{e} \rangle = 1.88 \pm 0.35$\,kpc, and absolute magnitudes in the range $-13.9 > M_{V} > -16.4$\,mag, being the mean value  $\langle M_{V} \rangle = -15.0 \pm 0.9$\,mag. They also have surface brightnesses $\mu_\mathrm{e} > 23$\,mag\,arcsec$^{-2}$, and a mean S\'ersic index of $\langle n \rangle = 1.44 \pm 0.53$. This means that the UDGs candidates have larger radii and fainter effective surface brightness than the ETG population, at the same absolute magnitude, as can be seen in Figs.\,\ref{fig:Mv_re} and\,\ref{fig:Mv_mue}, respectively. 
We will come back to this topic in the Discussion.

\subsection{Spatial distribution}\label{sec:spatial_distribution}
\begin{figure}
  \centering
      \includegraphics[width=0.95\columnwidth]{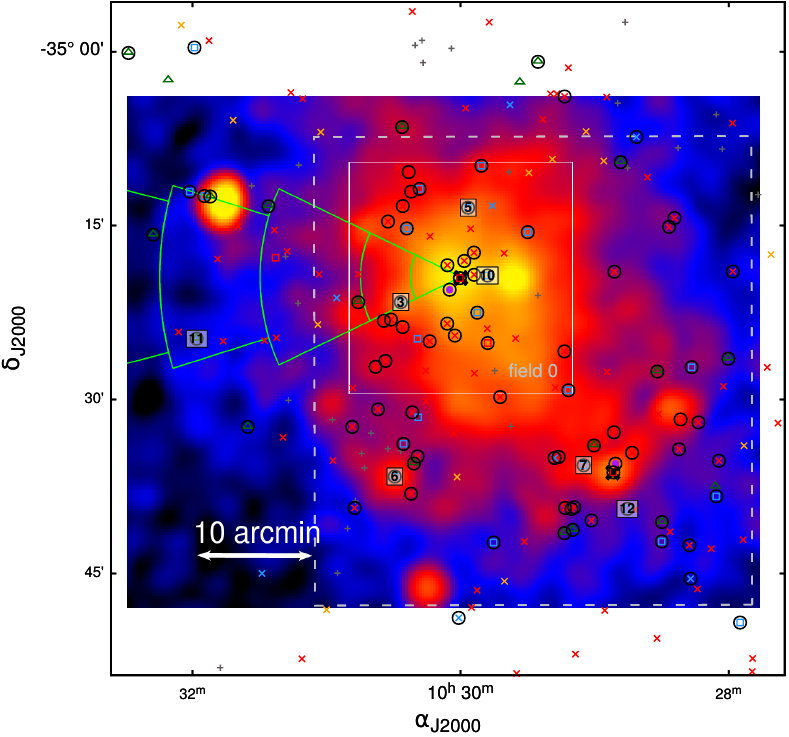}
      \caption{Central region of the Antlia cluster (field 0), with the location of the sample galaxies (symbols as in previous Figures) superimposed on an smoothed X-ray {\it ROSAT} PSPC (0.5--2\,keV band) map reproduced from Fig.\,1 (upper left) of  \citet{2016ApJ...829...49W} by permission of the AAS. X-ray intensity varies from high in yellow-red, to low in blue-black. The grey dashed line represents the MOSAIC field 0,  and the white square and pie regions correspond to different X-ray observations. Numbers identify bright galaxies listed in Table\,\ref{tab:topmagnitude}.}
      \label{fig:Wong-f1}
\end{figure}

\begin{figure*}
  \centering
  \includegraphics[width=0.85\textwidth]{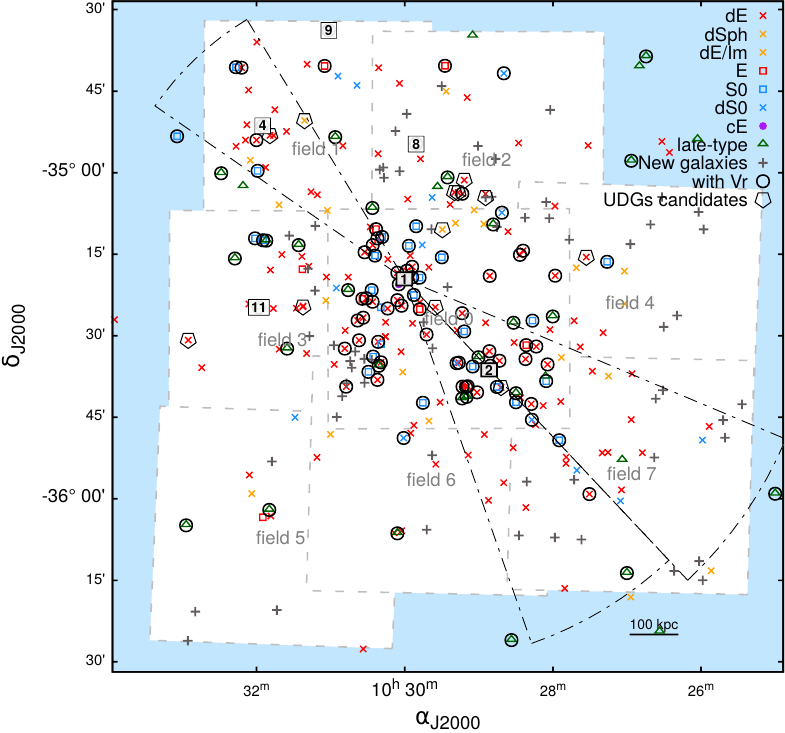}
  \caption{Composite image of the 8 MOSAIC\,II fields with Antlia galaxies from the complete sample superimposed. Symbols are explained on the upper right corner and numbers 4, 8, 9, and 11 identify galaxies brighter than $T_1 = 20$ (listed in Table\,\ref{tab:topmagnitude}), located outside field 0. Both dominant galaxies are also labelled. The three circular sectors indicate high density of galaxies (see  Fig.\,\ref{fig:hdistance_185_pie-1} and Sec.\,\ref{sec:spatial_distribution}).
  }\label{fig:radec1}
\end{figure*}

In order to study the projected spatial distribution, a position within the cluster should be taken as a reference point. Although the Antlia cluster has two dominant cD galaxies, to select such central point we will take into account the intensity of the X-ray emission, according to the map given by \citet[see their fig.\,1.]{2016ApJ...829...49W}. It can be seen that the X-ray intensity is clearly higher at the position of the galaxy NGC\,3268, while only an extension of lower intensity towards NGC\,3258 is present. Thus, in the following NGC\,3268 will be considered as the centre of the projected spatial distribution. In addition, NGC\,3268 is the centre of the youngest sub-cluster in Antlia \citep{2015MNRAS.452.1617H}. In Fig.\,\ref{fig:Wong-f1}, we present the position of the galaxies in our sample, superimposed to the X-ray map from \cite{2016ApJ...829...49W} that covers the central part of the cluster. The smoothed colour distribution shows the X-ray intensity (high: yellow - red, low: dark blue). The grey dashed line represents the boundaries of MOSAIC field 0, the symbols are as in the previous Figures, and the white square and pie regions correspond to different X-ray observations. We have also indicated the position of galaxies brighter than $M_{T_{1}} = -20$\,mag in Table\,\ref{tab:topmagnitude}. In this case, the locations of NGC\,3268 and NGC\,3258 are indicated with the upper and lower black crosses, respectively. Four of the brightest galaxies in our sample (FS90 172, 253, 300, and 304; or id 8, 9, 4, and 11, respectively) are located outside field 0, on field 1 (toward the north-east of field 0), showing that the brightest galaxies lie in the direction that connects both cD galaxies, coincident with the main axis of the elongated X-ray distribution (see Fig.\,\ref{fig:radec1}).

The Antlia cluster projected spatial distribution of the complete sample studied in this paper is shown in Fig.\,\ref{fig:radec1}, including the eight MOSAIC\,II fields (numbered from 0 to 7). Taking into account the overlap between fields, the total area covered in this paper is about 100\,arcmin $\times$ 100\,arcmin ($\sim$ 1000 $\times$ 1000\,kpc$^{2}$). Galaxy symbols are explained in the plot, numbers identify galaxies brighter than $T_{1} = 20$ located outside field 0, and the dominant galaxies are also highlighted. We clearly indicate galaxies with radial velocity measurements (encircled in black), new uncatalogued galaxies (grey crosses), and UDG candidates (black pentagons).
\begin{figure*}
  \begin{center}
    \subfigure[]{\label{fig:hdistance_185-1}%
        \includegraphics[width=0.50\textwidth]{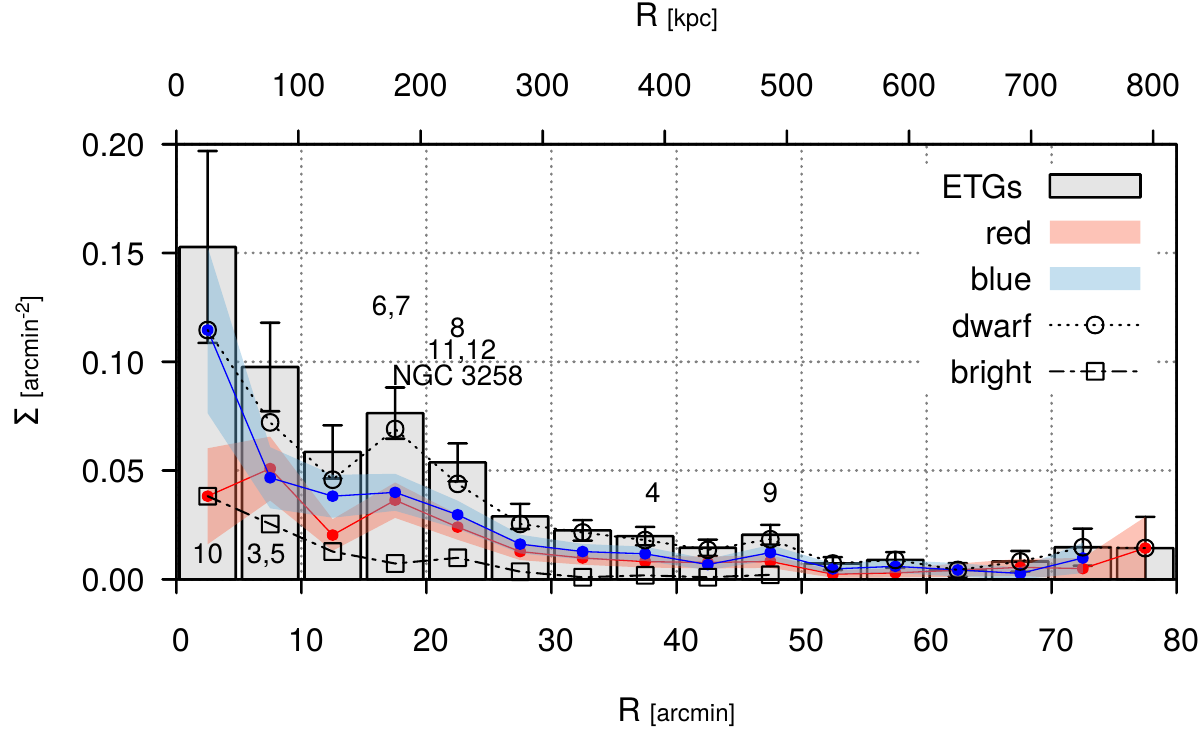}}%
    \subfigure[]{\label{fig:hdistance_185_pie-1}%
        \includegraphics[width=0.50\textwidth,scale=0.1]{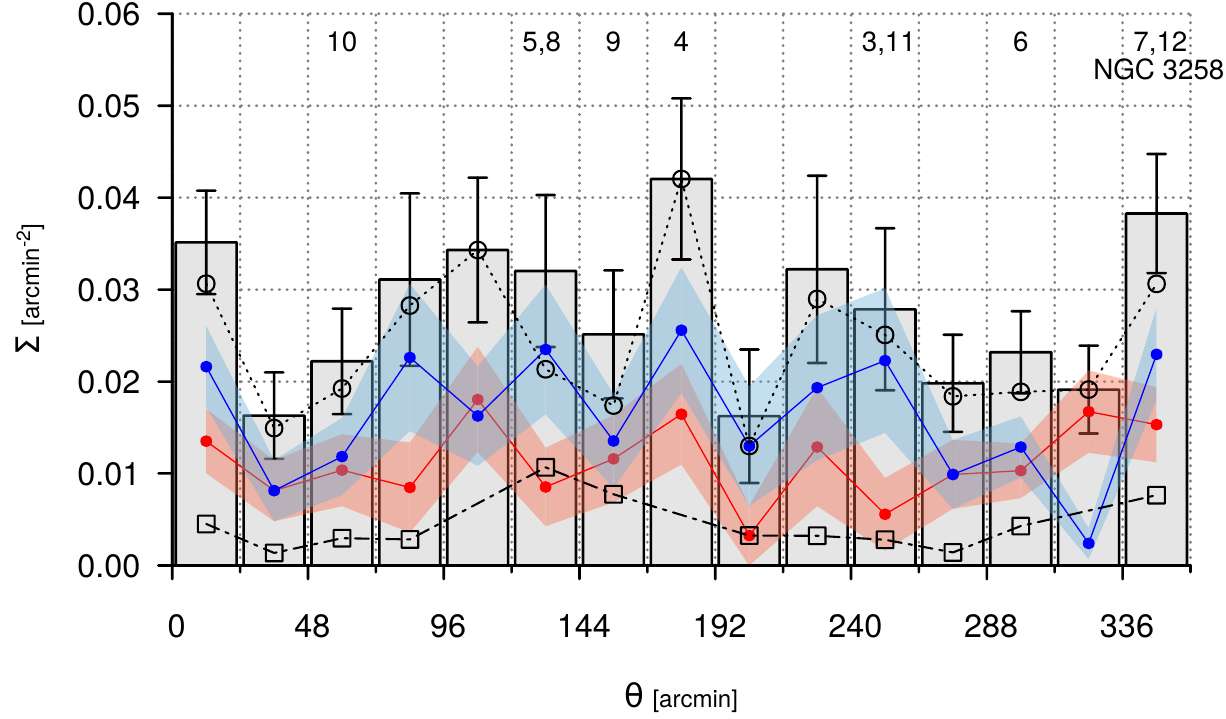}}%
    \caption{Projected number density of Antlia galaxies with respect to the centre of the distribution, taken at NGC\,3268, for all ETGs in the final sample. The galaxies brighter than $T_{1} = 20$\,mag are indicated in the corresponding bin. Blue and red lines correspond to galaxies bluer and redder than the CMR, while open squares and circles indicate bright and faint ETG subsamples, respectively. Panel\,(a): Density profile of the projected number density $\Sigma$ versus cluster-centric radius $R$.  Panel\,(b): Azimuthal distribution of the projected number density, taking the direction to NGC\,3258 as $\theta = 0^{\circ}$.}\label{fig:hdistance_all}
  \end{center}
\end{figure*}

\begin{table}\setlength{\tabcolsep}{2pt}
  \caption{Washington photometry of Antlia galaxies brighter than $M_{T_1} = -20$\,mag.  Radial velocities (last column) were taken from \protect\cite{2015A&A...584A.125C}.}\label{tab:topmagnitude}
  \begin{tabular}{cccccccc}
    \hline\rowcolor{white}
    id & FS90 & NGC/IC & $\alpha_{\mathrm{J}2000}$ & $\delta_{\mathrm{J}2000}$ & $(C-T_1)_{0}$ & $T_{1_0}$ & $V_{R}$ \\ 
      &      &        &      &          & 
    [mag]          & [mag]      & [\kms] \\
    \hline 
    1 & 185 & 3268 & 10:30:00.65 & -35:19:31.60 & 1.93 &  9.67 & 2800 $\pm$ 21 \\  \rowcolor{tableShade}
    2 & 111 & 3258 & 10:28:53.29 & -35:36:20.10 & 1.94 & 10.31 & 2792 $\pm$ 50 \\  
    3 & 224 & 3271 & 10:30:26.42 & -35:21:34.40 & 1.93 & 10.86 & 3737 $\pm$ 27 \\  \rowcolor{tableShade}
    4 & 300 & 3281 & 10:31:52.00 & -34:51:12.37 & 1.87 & 11.04 & 3200 $\pm$ 22 \\ 
    5 & 184 & 3269 & 10:29:57.03 & -35:13:27.57 & 1.82 & 11.58 & 3754 $\pm$ 33 \\  \rowcolor{tableShade}
    6 & 226 & 3273 & 10:30:29.09 & -35:36:36.60 & 2.00 & 11.68 & 2660 $\pm$ 08 \\
    7 & 125 & 3260 & 10:29:06.33 & -35:35:42.45 & 1.98 & 12.10 & 2439 $\pm$ 46 \\  \rowcolor{tableShade}
    8 & 172 & 2584 & 10:29:51.46 & -34:54:42.49 & 1.75 & 12.12 & 2549 $\pm$ 19 \\
    9 & 253 & 2587 & 10:30:59.84 & -34:33:44.52 & 1.96 & 12.21 & 2111 $\pm$ 18 \\  \rowcolor{tableShade}
    10 & 168 & 3267 & 10:29:48.58 & -35:19:19.03 & 1.79 & 12.23 & 3709 $\pm$ 33 \\
    11 & 304 & 3258D & 10:31:55.70 & -35:24:35.15 & 2.04 & 12.30 & 2476 $\pm$ 16 \\  \rowcolor{tableShade}
    12 & 105 & 3257 & 10:28:47.08 & -35:39:28.69 & 1.86 & 12.62 & 3237 $\pm$ 15 \\ 
    \hline
  \end{tabular}
\end{table}

In order to study any possible correlation between galaxy colour and projected spatial distribution, as well as for the following analysis, we split the ETG sample into galaxies bluer and redder than the CMR (Eq.\,\ref{eq:regression}). In Fig.\,\ref{fig:hdistance_185-1}, the global density profile shows that the main concentration of ETGs is clearly around NGC\,3268 (i.e. cluster-centric radius $R = 0$\,arcmin), while another over-density is present at the fourth annulus, close to NGC\,3258. The galaxies bluer than the CMR are more concentrated up to $R\sim 300$\,kpc, with a smooth decline. Particularly, in the innermost annulus, bluer ETGs more than double the redder ones. The over-density close to the location of NGC\,3258 is also detected in the galaxies redder than the CMR, and it corresponds to a younger sub-cluster that is probably infalling to NGC\,3268 \citep{2008MNRAS.386.2311S, 2015MNRAS.452.1617H}.

We also studied the projected spatial distribution considering giant (bright) and dwarf ETGs. The former, mostly Es and S0s, were represented in Fig.\,\ref{fig:hdistance_185-1} by empty squares, while the latter, mostly dEs, by empty circles. The adopted morphology for the FS90 galaxies was taken from their catalogue. The new galaxy candidates added to the final sample, mainly dwarfs, were selected by the elliptical shape on the images (see  \citetalias{2015MNRAS.451..791C}). Frequently, the distinction between dwarfs and giants is not clear. Rather than the magnitude, the central surface brightness is considered as a better option to separate dwarfs from giants \citep[e.g.][]{2009A&A...508..665B, 2009ApJS..182..216K}. We thus split the sample into dwarfs and giants, taking as limit between them: a central surface brightness $\mu_{0} = 19$\,mag/arcsec$^{2}$ ($V$-band), a magnitude $T_1 = 14$\,mag \citep{2008MNRAS.386.2311S}, or just the morphological classification. We then applied a KS test to compare their projected distributions (both versus cluster-centric radius or azimuthal angle), and all resulted comparatively quite similar, i.e. the KS test shows that we cannot distinguish statistically the different samples, when they are split by $\mu_{0}$, $T_{1}$ or the morphological classification. For the rest of the analysis, we will apply a dwarf/bright ETGs  separation using $\mu_{0}$.

The projected density of the faint ETGs surpasses that of the bright ones in every annulus of the profile (Fig.\,\ref{fig:hdistance_185-1}), even in the central region of the cluster, where the largest number of bright galaxies is found. To determine the principal direction along which the galaxies are distributed, we show the azimuthal projected distribution in Fig.\,\ref{fig:hdistance_185_pie-1}, which is also centred on NGC\,3268. It can be seen that the main over-densities take place in the direction of NGC\,3258 ($\theta = 0^{\circ}$) and in its exactly opposite direction ($\theta = 180^{\circ}$), i.e. an axis connecting both cD galaxies defines the direction along which galaxy density is enhanced. For the sake of clarity, the circular sectors corresponding to these over-densities are also shown in Fig.\,\ref{fig:radec1}. The correlation between the $X$ and $Y$ coordinates of the galaxies can be quantified with the Pearson coefficient \citep{Freedman2007} which is about 0.2, a moderate correlation with a confidence level greater than 95 per\,cent. Also, the orientation of the correlation ellipse  matches the direction of an axis joining NGC\,3258 and NGC\,3268. This indicates that we cannot reject the hypothesis that the projected distribution is elongated in the direction corresponding to the over-densities shown in Fig.\ref{fig:hdistance_185_pie-1}.
If we compare the projected distributions of dwarf against bright galaxies using a KS test, we obtain statistics of $0.277$ and $0.118$ ($1-p$ values of $0.9868$ and $0.2459$) for radial and azimuthal distributions, respectively. Such results indicate that we can reject the hypothesis that the two populations are drawn from the same distributions,  with a confidence level of 95 per cent, when we consider the radial distribution; but we cannot reject such hypothesis for the azimuthal distribution centred on NGC\,3268.

On the other hand, the radial and azimuthal distributions for the full final sample were also compared using the KS test, in order to look for any possible colour-projected spatial distribution trend.  In this case, the KS test gives statistics of  0.1072 and 0.1004 (with $1-p$-value of 0.5797 and 0.5144) for the radial and azimuthal  distributions, respectively. This means that we cannot reject the hypothesis that the ETGs redder and bluer than the CMR come from the same parent distribution.

\begin{figure}
  \centering
  \includegraphics[width=0.95\columnwidth]{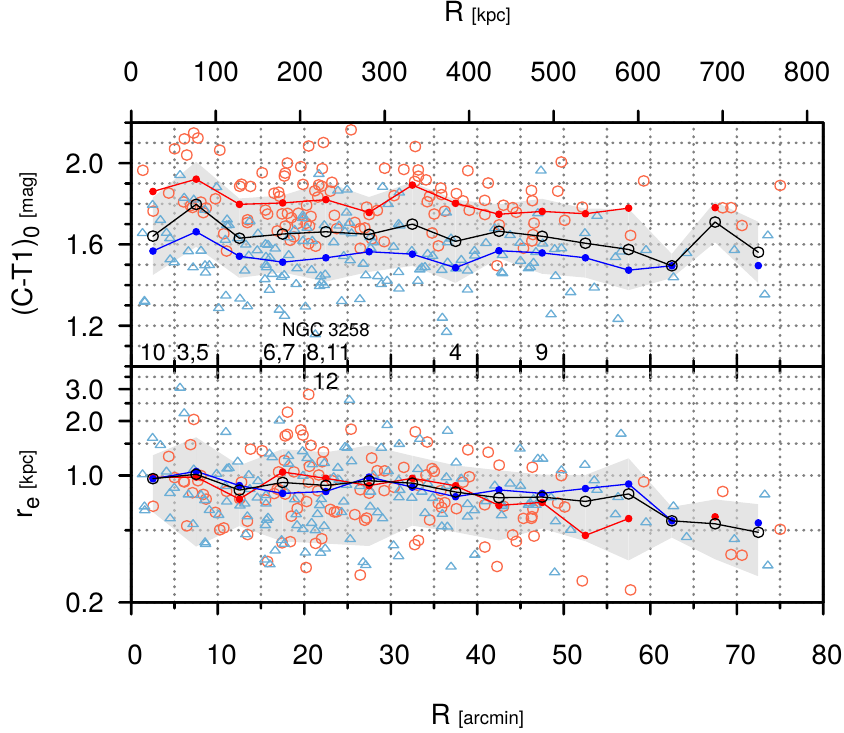}
  \caption{ETGs colour and effective radius as a function of cluster-centric distance, centred on NGC\,3268. Galaxies bluer and redder than the CMR, are identified with blue triangles and red circles, respectively. Filled blue and red circles correspond to the mean values in each bin. The mean values for all galaxies in each bin are shown with open black circles while their dispersion is represented by the grey band. Numbers between panels   identify the galaxies brighter than $T_{1}=20$\,mag  (Table\,\ref{tab:topmagnitude}) in the corresponding bin, as in previous figures.}\label{fig:h_185-CT1-re}
\end{figure}

In Fig.\,\ref{fig:h_185-CT1-re} we investigate the relation of the ETGs $(C-T_1)_{0}$ colour (top panel) and effective radius (bottom panel) versus the cluster-centric radius ($R$, centred on NGC\,3268). 
The galaxy sample is again divided into those bluer (blue triangles) and redder (red open circles) than the CMR regression, while filled circles show the respective mean values in each bin. Black open circles correspond to the mean values of all ETGs in each bin and the grey band indicates their dispersion. As a reference, the brighter galaxies have been identified in the respective bins, following the id given in Table\,\ref{tab:topmagnitude}. 
The colour of the whole final sample as well as the blue and red subsamples (top panel), do not show any clear gradient along the cluster-centric distance, but seem to be almost constant at $\langle (C-T_1)_{0} \rangle \sim 1.69 \pm 0.09$\,mag for the whole radial range (the correlation coefficient for the linear fit is $coor = 0.06$). One of the most populated bins corresponds to $R = 22.5$\,arcmin, which agrees with the position of NGC\,3258 and has the largest dispersion $\sigma_{22.5} = 0.24$\,mag. This behaviour is the same if we refer to the bluer or redder subsamples, separately.

Now, we consider the radial trend of effective radius (bottom panel) of the whole ETG sample, excepting the following outliers: three bright galaxies with $r_\mathrm{e} > 5$\,kpc (NGC\,3258, NGC\,3268, and FS90\,253) and the two cEs (FS90\,110 and FS90\,192). The mean effective radius close to the cluster centre (NGC\,3268) is $\langle r_\mathrm{e}\rangle = 0.9 \pm 0.2$\,kpc, which is mainly traced by the dEs \citep[e.g.][]{2008MNRAS.386.2311S}. A correlation ($coor = 0.9$ for a linear fit of $r_\mathrm{e}$ vs. $R$) is present along the cluster-centric distance, in the sense that the mean effective radius decreases towards the outer regions, even beyond the sub-group centred on NGC\,3258. The galaxies that present redder and bluer colours than the CMR, follow the same trend along the cluster radius. In particular, at $R = 22.5$\,arcmin, the redder galaxies have the largest dispersion in mean effective radius $\sigma_{22.5} = 0.57$\,kpc, while if we consider the whole sample or the bluer galaxies, the largest dispersion is present at the next bin, at $R = 27.5$\,arcmin with $\sigma_{27.5} = 0.52$ and $0.61$\,kpc, respectively.

\subsection{Radial velocity distribution}
\begin{figure*}
  \begin{center}
    \subfigure[]{\label{fig:vrhistogram-1}%
      \includegraphics[width=0.49\textwidth]{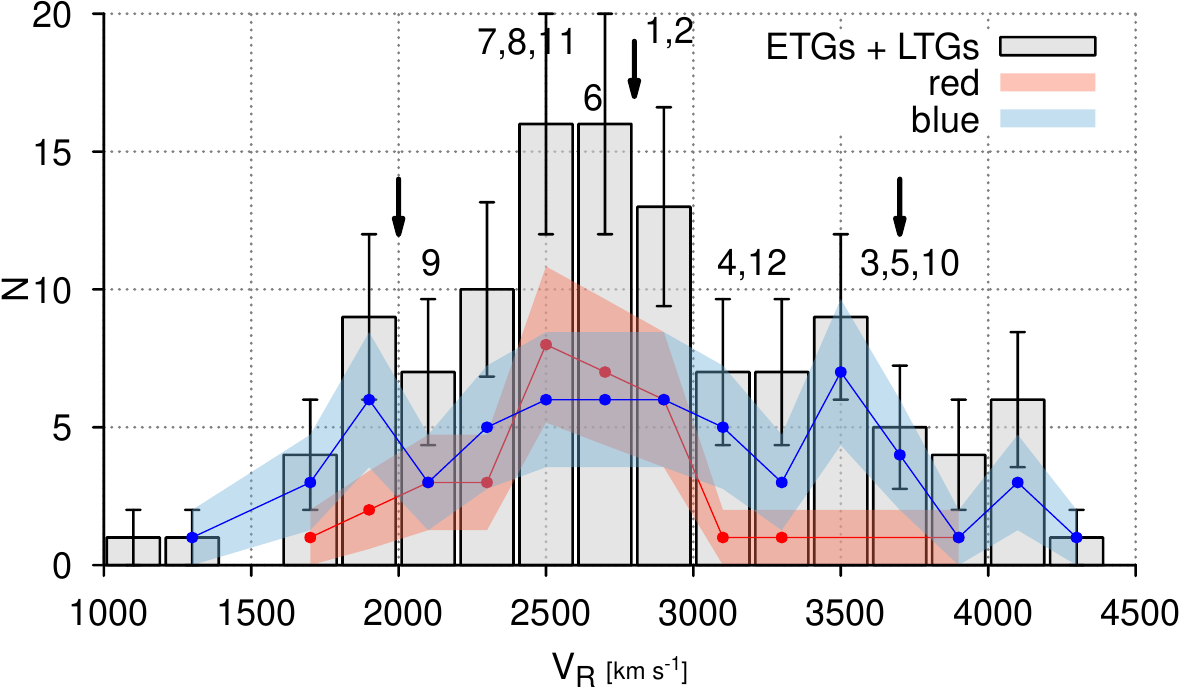}}
    \subfigure[]{\label{fig:vrmap}%
      \includegraphics[width=0.49\textwidth]{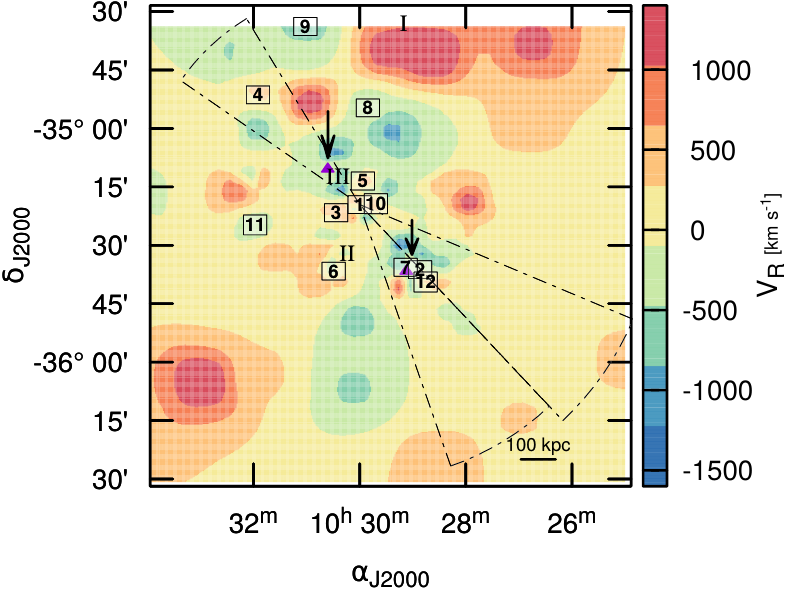}}\\
    \subfigure[]{\label{fig:radec-vrhistogram-mem001020R}%
     \includegraphics[width=0.49\textwidth]{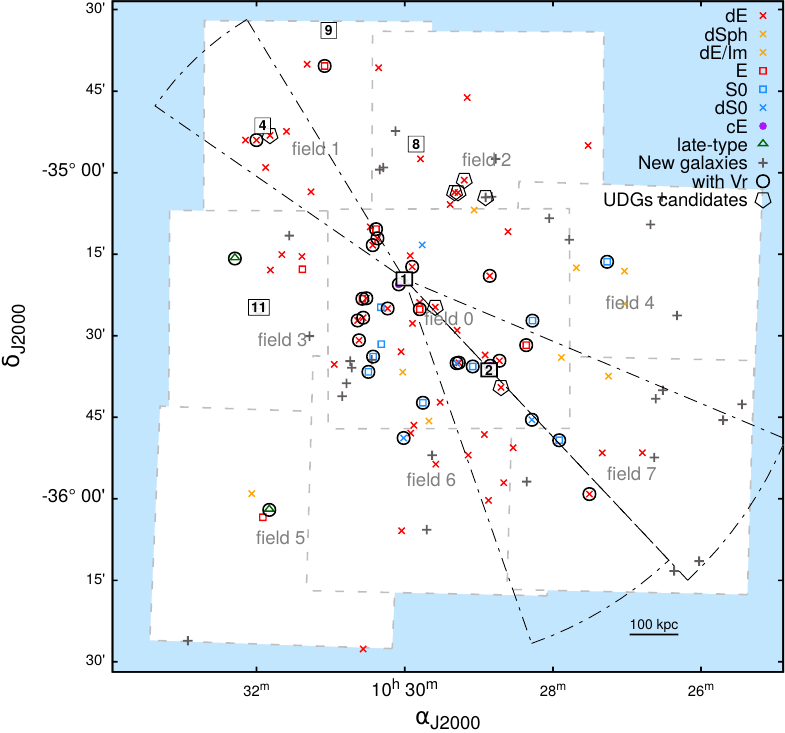}}
    \subfigure[]{\label{fig:radec-vrhistogram-mem001020B}%
      \includegraphics[width=0.49\textwidth]{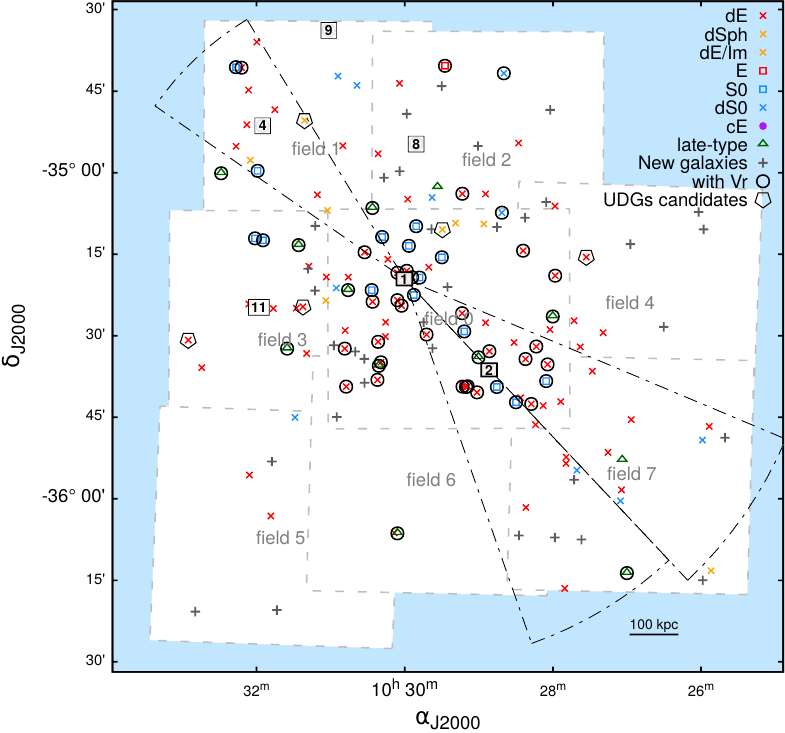}}
    \caption{(a) Radial velocity histogram of all galaxies in the sample split into redder and bluer than the CMR. With black arrows we indicate the main peaks in the histogram obtained by \protect\cite{2015A&A...584A.125C}, and identify the brightest galaxies in the respective bins, following the id given in Table\,\protect\ref{tab:topmagnitude}. (b) The Antlia cluster radial velocity map and the relation with the projected spatial distribution. With filled triangles (highlighted with arrows), we shown the position of the two centroids (see text). (c) and (d) The projected spatial distribution of redder and bluer galaxies than the CMR are shown, respectively. We also show on the maps the three circular sectors with the highest galaxy densities.}\label{fig:vr-color}
  \end{center}
\end{figure*}

In this Section we revisit the radial velocity distribution as a  function of $R$. 
The radial velocities of galaxies in the Antlia Cluster core were extensively studied by \cite{2015A&A...584A.125C}. We can now complement them with our photometry and structural parameters. In Fig.\,\ref{fig:vrhistogram-1} we show the radial velocity histogram of the total number of galaxies with radial velocities lying within the membership range (grey), and split the sample into galaxies redder and bluer than the CMR. The three arrows indicate the locations of the main peaks in the histogram obtained by \cite{2015A&A...584A.125C}: $2060 \pm 200$\,\kms, $2780 \pm 100$\,\kms, and $3600 \pm 130$\,\kms. It can be seen that the extreme radial velocity peaks seem to be related with the bluer ETGs, while red ETGs tend to be concentrated to the central peak (which includes the brightest galaxies) and whose location could be related with the X-ray major emission. On the other hand, the bluer galaxies tend to display extreme radial velocities within the membership range, and have a flat projected distribution around the central field.
Regarding the radial velocity distribution in this central region, we note that there are two S0 galaxies, NGC\,3273 and NGC\,3260 (FS90\,226 and FS90\,125, respectively), located in areas of high X-ray emission between both cDs, that have the lowest radial velocities within the central field. On the other side, the highest radial velocities in this field correspond to NGC\,3269 and NGC\,3271 (FS90\,184 and FS90\,224), that lie near NGC\,3268 in the area of maximum X-ray emission of the cluster.

In Fig.~\ref{fig:vrmap} we represent the radial velocity distribution together with the spatial projected distribution of galaxies as a `heatmap', i.e. a two dimensional graphical representation that uses a colour palette to indicate the different velocity ranges, each one calculated as the distance weighted mean. We have decided to re-centre the $V_{R}$ range relative to the NGC\,3268 velocity, which is $\sim 2800$\kms. In this way, the yellow/green areas have a predominance of galaxies with similar $V_{R}$, red/orange and blue areas indicate extreme $V_{R}$ (within the membership range). We also indicate the position of the brightest galaxies as spatial reference of the substructures in the cluster (ids in Table\,\ref{tab:topmagnitude}). The three kinematic substructures studied by \cite{2015MNRAS.452.1617H}: I, II, and III, are also indicated. These structures were obtained taking into account the velocity dispersion referred to a local mean velocity. Additionally, we performed a spatial clustering study through the k-means numerical method \citep{1056489}. This algorithm is based on an Euclidean distance and iteratively calculates the distance between groups of objects, built from the average distance of each group. As a result, we can determine the centroid of each data group in the space of coordinates (in our case, the spatial projected coordinates). We show in Fig.~\ref{fig:vrmap}, with filled triangles (pointed at by arrows), the position of the centroids when two groups are considered as substructures, taking into account the members of the sample with $V_{R}$. Finally, in Fig.\,\ref{fig:radec-vrhistogram-mem001020R} and Fig.\,\ref{fig:radec-vrhistogram-mem001020B} we show separately the positions of the galaxies redder and bluer than the CMR.

\subsection{Luminosity function}\label{sec:LF}
\begin{figure}
  \centering
  \includegraphics[width=0.95\columnwidth]{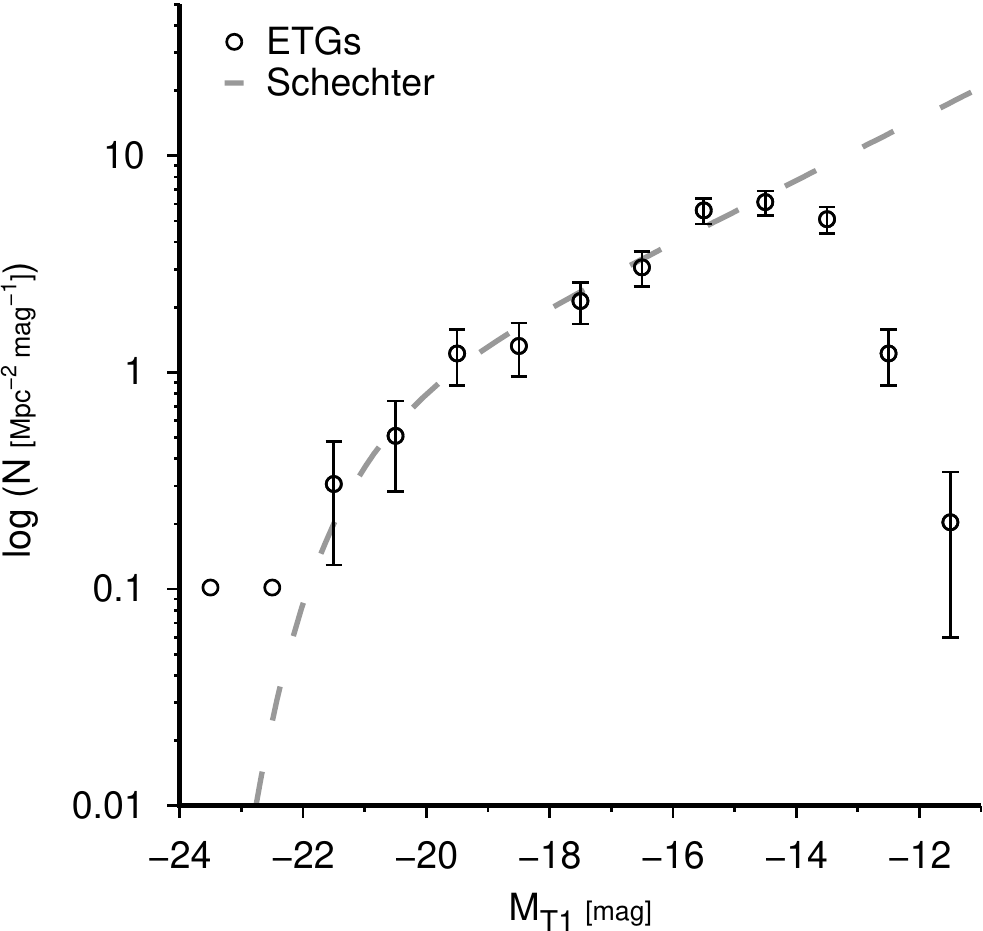}
  \caption{Luminosity function of the ETGs in the sample, and a Schechter function fit on dashed line.}\label{fig:funcion_de_luminosidad}
\end{figure}

The luminosity function (LF) of a cluster can be built by counting the number of galaxies in different magnitude bins. We show in Fig.\,\ref{fig:funcion_de_luminosidad} the LF, in the $T_1$-band, of the ETGs in the Antlia Cluster. We used the \memCDVRB ETGs in the sample, which are distributed on the eight MOSAIC\,II fields, covering an effective area of about 2.6\,deg$^{2}$ or 1\,Mpc$^{2}$ at the distance of 35.2\,Mpc \citep[$H_{0}=75$\kms\,Mpc$^{-1}$,][]{2003A&A...408..929D}. We used the Schechter function \citep{1976ApJ...203..297S},
\begin{equation}
  \varphi(M) dM = \varphi^{*} 10^{-0.4(M-M^{*})(\alpha +1)} \exp^{-10^{-0.4(M-M^{*})}} dM\textnormal{,}
\end{equation}
where $\varphi^{*}$ is the characteristic density, $M^{*}$ is the absolute magnitude at the slope break, and $\alpha$ is the slope at the faint luminosity end. We show in dashed-black line a fit of the LF in the magnitude range $-22 < M_{T_1} < -14$\,mag, with parameters: $\varphi^{*} = 0.55 \pm 0.36$, $\alpha = -1.37 \pm 0.03$, and $M^{*} = -21.70 \pm 1.36$, and $\upchi^{2} = 0.11$.

Due to the scarcity of bright (giant) ellipticals in the cluster, the Schechter fit cannot be well constrained at the bright magnitude side. The faint end is affected by the sample incompleteness. Since the sample was visually selected, automatic methods to calculate the completeness cannot be applied \citep{2012A&A...538A..69L}. To estimate the incompleteness, we used the structural characteristics of the faintest galaxies in the sample, determining the lowest effective surface brightness detected for a typical galaxy,  which turned out to be about 2 per cent at $M_{V} \sim -13$\,mag \citepalias{2018MNRAS.477.1760C}. The absence of a dip in the LF of Fig.\,\ref{fig:funcion_de_luminosidad}, which was extensively studied in different clusters and environments \citep[ and references therein]{2016ApJ...822...92L}, agrees with similar results in Fornax \citep{2003A&A...397L...9H, 2007ApJS..169..213J}, and the NGC\,5044 group \citep{2012MNRAS.420.3427B}.

\subsection{Estimation of stellar masses}\label{sec:stellar_masses}
In order to estimate the galaxy stellar mass of each object in the sample, we used the galaxy models presented in \cite{2005MNRAS.361..725B}, that provide the appropriate mass-luminosity ratio for different galaxy morphologies. The galaxy luminosity in the $T_1$-band was transformed into $V$ using the expression given in \citet[Table A1 and A2]{2005MNRAS.361..725B} and the $M/L$ ratios listed in \citet[Table 4]{2012MNRAS.420.3427B}.

\begin{figure}
  \centering
  \includegraphics[width=0.95\columnwidth]{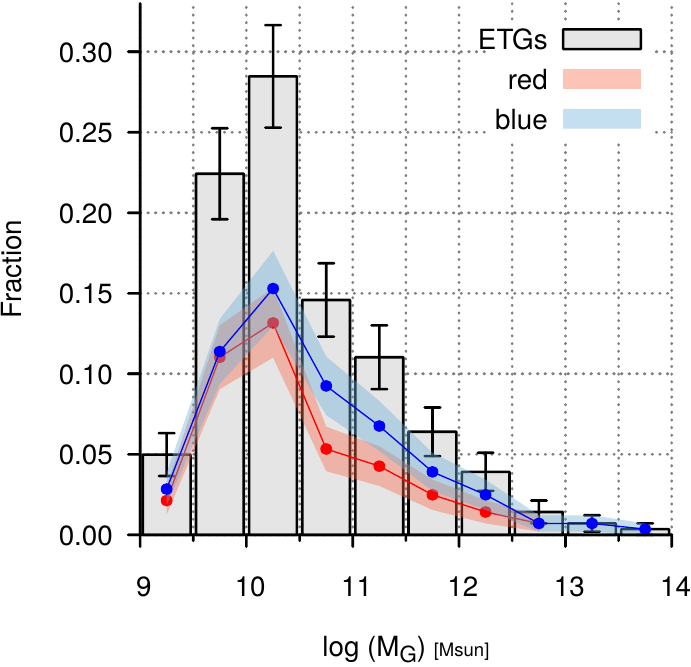}
\caption{Galaxy stellar mass histogram.}\label{fig:hmass-crop}
\end{figure}

The galaxy mass ($M_{\mathcal{G}}$) was estimated as follows:
\begin{equation}
M_{\mathcal{G}} = \frac{M}{L}\,\left[\frac{M_{\odot}}{L_{\odot}}\right] ~ 10^{-0.4\,(V - V_{\odot})}.
\end{equation}
In Fig.\,\ref{fig:hmass-crop} we show a stellar mass histogram for the \memCDVRB ETGs in the sample. All the galaxies span the range from $7.5 \lesssim \log(M_{\mathcal{G}})/\Msun \lesssim 14.0 $, with an abrupt cut at the lower mass limit. The logarithmic mean mass is $10.48 \pm 0.78$\,\Msun, which corresponds to $\langle M_{\mathcal{G}} \rangle = 3.0 \times 10^{10}$\,\Msun. We also show the distribution of blue/red galaxies (with respect to the CMR) in the same Figure. Bluer galaxies tend to be more frequent along almost all the mass range, while for intermediate masses ($\log(M_{\mathcal{G}})/\Msun \sim 10.0 - 11.5$), redder galaxies show a deficit with respect to the general trend.

\section{Discussion}\label{sec:discussion}
In this work, we study the largest ETG sample of the Antlia Cluster. Our main goals are to quantify the characteristics of the ETG population and to describe their projected spatial distribution as well as the correlations between their structural parameters. Table\,\ref{tab:new_antlia} consists of \totalcatalogue galaxies from all types, including those identified by \citetalias{1990AJ....100....1F}, \cite{2012MNRAS.419.2472S}, and \citetalias{2015MNRAS.451..791C}. Each source is clearly identified in the table. 

\subsection{Structural parameters}
\begin{figure*}
  \begin{center}
    \subfigure[]{\label{fig:hMv}%
      \includegraphics[width=0.25\textwidth]{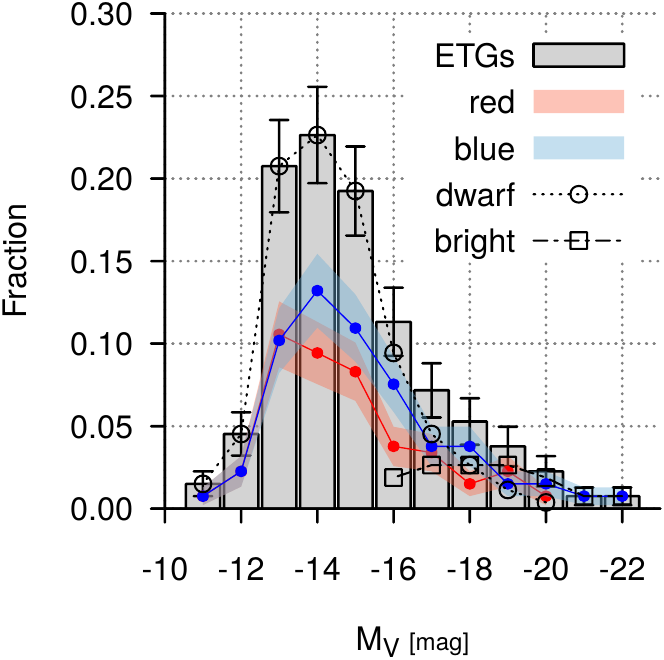}}%
    \subfigure[]{\label{fig:hre}%
      \includegraphics[width=0.25\textwidth]{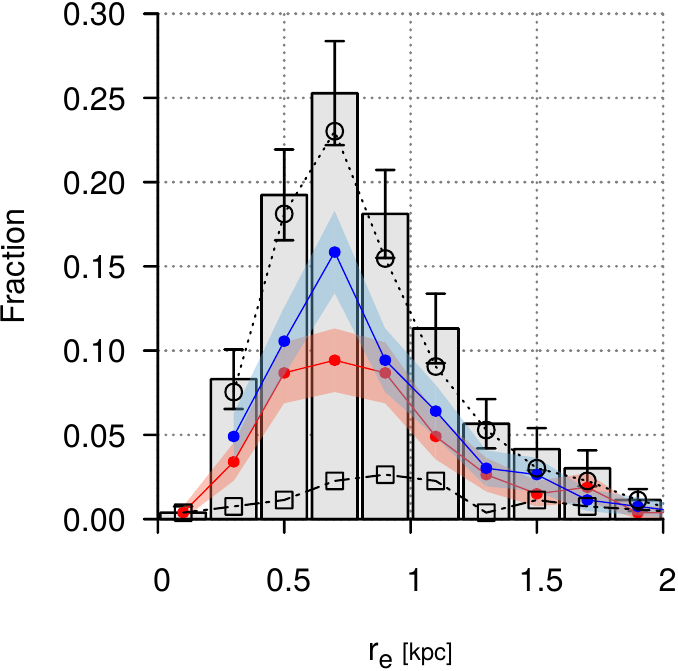}}%
    \subfigure[]{\label{fig:hn}%
      \includegraphics[width=0.25\textwidth]{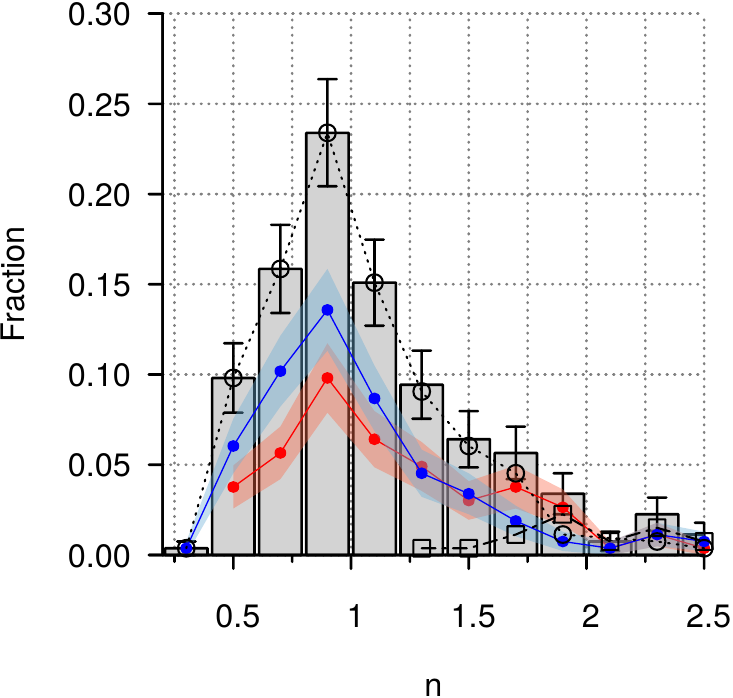}}%
    \subfigure[]{\label{fig:hmue}%
      \includegraphics[width=0.25\textwidth]{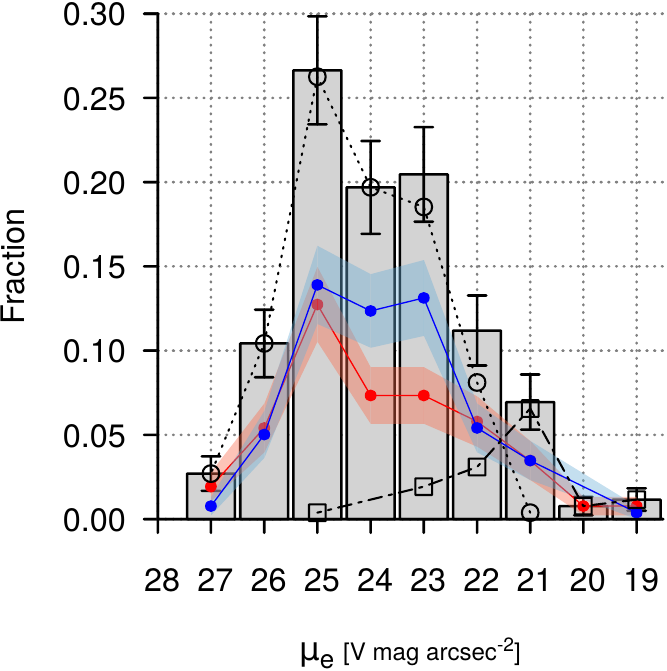}}
    \caption{Absolute magnitude and structural parameters distributions (see text).}\label{fig:histograms}
  \end{center}
\end{figure*}

Fig.\,\ref{fig:scaling_relations} presented the relations between the structural parameters of all \memCDVRB ETGs in the sample. Each parameter: $r_\mathrm{e}$, S\'ersic index $n$ and $\mu_\mathrm{e}$, was obtained from the fit of the respective surface brightness profile; their relationships reveal the different populations that coexist in the cluster. Now, in Fig.\,\ref{fig:histograms} we show the distributions of each parameter along with that of the total absolute magnitude $M_{V}$ \citepalias[for the maths, see][]{2018MNRAS.477.1760C}. On each graph, we show in grey the histogram of the ETGs in the sample, in red and blue the corresponding sub-populations with respect to the CMR, and we additionally show them separated into giant (empty squares) and dwarf (empty circles) ETGs, using as limit the central surface brightness $\mu_{0} = 19$\,\mags (on $V$-band). Fig.\,\ref{fig:hMv} shows the distribution of the total absolute magnitude, spanning a range of $-23 < M_{V} < -10$\,mag. The LSB galaxies dominate the faint magnitude range, with a maximum at $M_{V} \approx -14$\,mag. This behaviour is followed by both the red/blue galaxies, with the exception that the bluer galaxies seem to dominate in number at the faintest magnitudes. Fig.\,\ref{fig:hre}, shows a unique peak at $r_\mathrm{e} \lesssim 1$\,kpc, shared by the blue/red and dwarf/giant sub-populations. While small radii bins are populated mainly by dwarf galaxies, giant ones span a broad range in radius. We remind here that we use a one component fit to model the surface brightness profile, which is very accurate for most of the sample galaxies. A sudden drop can be seen at $r_\mathrm{e} \approx 1.5$\,kpc, which sets the radius where LSB galaxies start to be dominated by UDGs, as those found in Fornax \citep{2017A&A...608A.142V, 2018ApJ...855..142E}. If we take into account the dwarf population, the characteristic effective radius is $r_\mathrm{e} = 0.8 \pm 0.4$\,kpc ($T_{1}$-band), which is larger than the value found for the Fornax Cluster which, in fact, has a deeper photometry on the most recent surveys. The distribution of the S\'ersic index is shown in Fig.\,\ref{fig:hn}. The different behaviours for giant and LSB galaxies are evident, pointing to the different morphologies which shape their surface brightness profiles. The histogram has also a unique peak, for the whole sample. The mean value for LSB galaxies is $\langle N\rangle = \langle 1/n \rangle = 1.03 \pm 0.45$, which is also larger than the mean S\'ersic index found in different environments \citep{2009MNRAS.393..798D,2018ApJ...855..142E}. The last histogram, Fig.\,\ref{fig:hmue}, shows the distribution of $\mu_\mathrm{e}$. There is a main peak at $\mu_\mathrm{e} = 25$\,\mags, dominated by LSB galaxies, and a secondary peak at $\mu_\mathrm{e} = 23$\,\mags, which seems to lie on the bridge between LSBs and giant galaxies. The distribution of giant galaxies, in turn, has a main concentration at higher surface brightness, at $\mu_{e} = 21$\,\mags. On the other hand, both redder and bluer galaxies seem to follow the general distribution in effective surface brightness.

\subsection{Spatial distribution}
The projected spatial distribution of galaxies in the Antlia Cluster has been studied by \citetalias{1990AJ....100....1F}, using photographic plates with a limited magnitude range, but with an interesting dynamical range, which allowed them to define a very precise membership classification based on the galaxies morphology. The elongated distribution of the galaxy population was first suggested by \citetalias{1990AJ....100....1F}, using the brightest ETGs of the cluster. More recently, X-ray observations reinforced the idea of an axial distribution, by mapping the diffuse emission that connects NGC\,3258 with NGC\,3268 \citep{1997ApJ...485L..17P, 2000PASJ...52..623N}. Then, \cite{2016ApJ...829...49W} studied the cluster core (a bit larger than the MOSAIC\,II field 0, see Fig.\,\ref{fig:Wong-f1}) at 0.5-2\,keV and confirmed that the major emission of the X-ray map is centred at NGC\,3268 and is elongated on the direction of NGC\,3258 \citep{2008MNRAS.386.2311S, 2015MNRAS.452.1617H}. The latter dominates a younger galaxy subgroup that has recently merged with the first one. Additionally, \cite{2015MNRAS.452.1617H} report that the Compton-thick Seyfert\,II NGC\,3281 (id 4 on Table\,\ref{tab:topmagnitude}), has an associated H\textsc{i} absorption region. This galaxy lies ---in field 1-- in the direction that connects NGC\,3258 with NGC\,3268 (see Fig.\,\ref{fig:radec1}). Now, with a sample of \memCDVRB ETGs covering a magnitude range of $-10 < M_{V} < -22$\,mag, distributed around a cluster-centric distance $R = 700$\,kpc, we have built the projected number density  profile (Fig.\,\ref{fig:hdistance_all}). It can be seen that the spatial projected distribution of the ETGs has a main concentration around NGC\,3268 and a second concentration around 200\,kpc from the cluster centre, close to NGC\,3258. Along with this, the angular distribution shows that the ETGs are concentrated along a direction that connects both cD galaxies. On the other hand, redder and bluer ETGs seem to have peculiar characteristics that could be related with the evolutionary state of the cluster, the dynamical relation between the NGC\,3258 group and the cluster core, and/or the characteristics of the DM haloes of the cluster itself.

Taking into account the central area of the cluster, up to $R = 300$\,kpc, bluer ETGs seem to be more frequent, i.e. on the central field there are more bluer ETGs than redder ones. Within this area, the density profiles of redder and bluer ETGs fall to 0.012 and 0.010\,arcmin$^{-2}$, respectively. While the density of bluer ETGs is strictly decreasing along the whole radial range, redder ETGs display two overdensities near each cD. If this is not a consequence of a detection deficiency, it could be related to the ram-pressure stripping \citep[see e.\,g.][]{1972ApJ...176....1G} suffered by galaxies within the densest areas of the cluster, which would not happen if they were not linked gravitationally or immersed on the same DM halo \citep{2016A&A...591A..51S, 2019MNRAS.483.2251Z}. Another interesting feature is that the dwarf-to-giant ratio has a secondary peak near NGC\,3258 (at $R \sim 200$\,kpc), matching the high density X-ray emission centred on NGC\,3268. Since the Antlia cluster presents a very complex structure, involving a massive cD, associated with a high X-ray emission, and an accreted group of galaxies centred on another cD, it is possible that different physical processes are acting in different degrees. \cite{2011MNRAS.416.1197W} compare the nearest galaxy clusters spatial distributions and colours to semi-analytic models (SAM) on the low luminosity range of dEs. They collected several samples from the literature of the Virgo, Fornax, Coma and Perseus clusters, and carefully match them with different SAM samples. The physical properties derived from the SAMs seem to be in good agreement with the observed clusters, even Fornax and Virgo, that have been suggested to be in a non-virialized state. The only parameter that is not well reproduced by the SAMs is the dwarf-to-giant ratio, which gives too high values. They also compared the observed radial density profiles with the simulated average profiles (their fig.\,4.). The dwarf ETGs on Antlia cover a magnitude range between $-19 < M_{V} < -11.1$\,mag, which implies approximately a number density of $\log(\Sigma) \approx 2.5$\,Mpc$^{-2}$ between the central cluster galaxy (NGC\,3268) and NGC\,3258 ($\sim 0.2$\,Mpc), falling down to $\log(\Sigma) \approx 1.7$\,Mpc$^{-2}$, around 0.7\,Mpc from the centre. The flattening number of dwarfs around the central area of the cluster, together with the central excess of bluer dwarfs, may be indications of a dynamically young population, as was suggested for the Virgo cluster. This may be produced by stronger tidal disruption suffered by the faintest galaxies in the densest area (which is consistent with the broad velocity distribution, shown in Fig.\,\ref{fig:vrhistogram-1}).

Any correlation between galaxy colours and spatial distribution arises from the colour-density or morphology-density relations \citep{1980ApJ...236..351D}, which states that there is a correlation between morphological types and the environments in which the galaxies evolve. Many studies confirm that while dEs are commonly found in the centre of clusters, star-forming dwarf irregulars are preferentially located in low density environments such as cluster outskirts. A more detailed view can be found, for example, in \cite{2009A&A...508..665B}, who study the population of dEs in the multiple-cluster system Abell 901/902 (at $z = 0.165$), between $-16 > M_{r} > -26$\,mag. They find that the nucleated dwarf elliptical (dEn) galaxies in their sample have different properties compared to the non-nucleated dE: they are more compact, more concentrated to the clusters centres, and they are rounder. These results seem to be largely due to tidal and ram pressure stripping \citep{2003AJ....125.1926G}, and even the preprocessing in galaxy groups (like the NGC\,3258 sub-cluster in Antlia) have strongly affected the dwarf population. They also found that the colours of dEs depend on their location within the cluster, in the sense that redder dEs are concentrated to the cluster centre. This behaviour is exactly opposite to our findings in Antlia, although, as seen in Fig.\,\ref{fig:hdistance_185-1}, the effect is mostly restricted to the innermost bin, containing a low number of galaxies. Another different feature in the Abell 901/902 system is its small colour gradient, as already reported in other galaxy clusters and groups. In Fornax, \cite{2017A&A...608A.142V} speculated that the galaxies that have spent enough time in the core of the cluster, should suffer the removal of their cold gas by ram pressure stripping. So, this could affect the colour of the central galaxies. The constant colour distribution along the Antlia cluster, until 60\,arcmin ($\sim 600$\,kpc), may indicate that Antlia is on a different stage on its evolutionary process.

\subsection{Radial velocity distribution}
The substructures of the Antlia Cluster have also been studied by \cite{2015MNRAS.452.1617H} in relation to H{\sc i} and star formation regions. Based on published radial velocities, they measured the relative kinematic deviation from the mean velocity that represents the entire cluster. Besides the obvious structures associated to NGC\,3268 (which was stated as cluster centre) and NGC\,3258, they proposed three more kinematic structures that coexist in the cluster \citep[see their fig. 7]{2015MNRAS.452.1617H}, which could be analysed together with our Fig.\,\ref{fig:vr-color}: the substructure {\it I} (at coordinates 10:29:12.00, -34:30:00) lies in Field\,2, on the cluster outskirts. Within this area, the most extreme radial velocity galaxies, i.e. FS90\,152, FS90\,101 are found. The substructure {\it II} (at 10:30:24.00, -35:36:00) is located at the cluster centre, and is associated with FS90\,226 (id\,6) that has a radial velocity representative of the brightest members. The substructure {\it III} (at 10:30:24.00, -35:12:00), located in Field\,0, is aligned with the direction connecting NGC\,3258 and NGC\,3268, and agrees with the position of one of the centroids calculated with the k-means method.

The radial velocity histogram (Fig.\,\ref{fig:vrhistogram-1}), shows that the bluer galaxies, that numerically dominate the innermost radial bin, have a broad range of velocities (between the membership limits) which may be interpreted as a result of the environment effects produced by the merger process blurring out the substructures in the area of the massive galaxies \citep[see also][]{2015MNRAS.452.1617H}. A different possible explanation was suggested by \cite{2015A&A...584A.125C}, using just a sample of radial velocities.

\subsection{Luminosity function}
The Antlia LF, which was described in Sec.\,\ref{sec:LF} with a Schechter fit: $\alpha = -1.37\pm0.03$, and $M^{*} = -21.70\pm1.36$, can be compared with those obtained in Fornax \citep{2019A&A...625A.143V} and Virgo clusters \citep{2016ApJ...824...10F}: $\alpha = -1.31\pm0.07$ and $\alpha = -1.34^{+0.017}_{-0.016}$, respectively. It is important to note that the Virgo and Fornax values are completeness corrected and obtained from deeper images than ours in the present paper. Other different environment galaxy systems are Hydra\,I \citep{2008A&A...486..697M} and Centaurus \citep{2009A&A...496..683M} clusters, that have shallower faint-end slopes: -1.13$\pm$0.04 and -1.08$\pm$0.03, respectively. These values are in agreement with the LF slope of the Fornax cluster centre, when dEs are considered: $-1.09\pm0.10$ \citep{2019A&A...625A.143V}. 

The LF shape is closely related with galaxy mergers, that are more frequent in galaxy groups because they can change the luminosity distribution of the galaxies involved \citep{2004MNRAS.355..785M}. In this sense, the presence of a dip in the LF of high-velocity dispersion clusters can be related with mergers \citep{2016ApJ...822...92L}, which would not be the case of the Antlia cluster, given its smaller velocity dispersion \citep[$\sigma_v= 500$\,\kms;][]{2015MNRAS.452.1617H}.

\subsection{UDG candidates}
On the LSB regime, between $-17.0 < M_{V} < -13.5$\,mag \citep{2017A&A...608A.142V}, we found 12 galaxies that present $r_\mathrm{e} > 1.5$\,kpc and $\mu_{0} > 25.0$\,\mags. These are the characteristics of the so called UDGs which were found in several clusters and groups \citep{2015ApJ...809L..21M, 2015ApJ...798L..45V, 2015ApJ...813L..15M, 2017MNRAS.468.4039R}. In Fig.\,\ref{fig:Mv_re}, they can be identified as a group of galaxies (indicated as empty pentagons) extending to higher $r_\mathrm{e}$ values; their location has also been highlighted with a green shading to clearly show the position of the Antlia UDGs. These galaxies were also shown in Fig.\,\ref{fig:Mv_mue}, where it can be seen that their $\mu_\mathrm{e}$ depart from the general trend, reaching almost 27\,\mags.  We show in Table\,\ref{tab:udg-candidates} the mean values of the UDGs structural and photometric parameters, separating them into two groups: {\it i)} {\it Group\,1} (G1) is composed by FS90\,135, 268, 270, 329 and {\it ii)} {\it Group\,2} (G2) by ANTLJ102936-352445, ANTLJ102843-353933, ANTLJ102856-350435.5, FS90\,52, 141, 143, 154 and 293. These two groups clearly differ in mean magnitude: $M_{V} \sim -15.98 \pm 0.28$\,mag, and $M_{V} \sim -14.58 \pm 0.28$, respectively, but they share almost the same mean $r_\mathrm{e} \sim 1.8 - 2.0$\,kpc. We performed a principal component analysis (PCA) of the UDGs parameters that also confirmed the segregation into two groups.

\begin{table}\setlength{\tabcolsep}{2pt}
  \caption{Antlia UDG candidates}
\label{tab:udg-candidates}
\begin{tabular}{ccccccc}
\hline
              & $C-T_1$       & $r_\mathrm{e}$       & $\mu_\mathrm{e}$      & $n$  & $R$ \\
              & [mag]         & [kpc]                &
                [\mags]             &      & [kpc] \\            
\hline
{\it Group 1} & 1.58$\pm$0.34 & 2.03$\pm$0.54 & 24.51$\pm$0.27 & 1.29$\pm$0.38 & 33.35$\pm$9.51\\ \rowcolor{tableShade}
\it {Group 2} & 1.81$\pm$0.30 & 1.81$\pm$0.22 & 26.53$\pm$0.42 & 1.51$\pm$0.60 & 19.85$\pm$9.07\\ 
\hline
\end{tabular}
\end{table} 

For comparison, we added in Figs.\,\ref{fig:Mv_mue} and \ref{fig:Mv_re}, the Fornax sample studied by \citet[which includes LSBs and UDGs]{2017A&A...608A.142V} that spans a range of $r_\mathrm{e}$ between 1.6 and 11.3\,kpc, largely exceeding the range of the Antlia ones. Larger values of $r_\mathrm{e}$ are also found in Virgo \citep{2015ApJ...809L..21M} and Coma \citep{2015ApJ...798L..45V}, which could indicate that the particular Antlia  evolutionary state, taking into account the ongoing merger with the NGC\,3258 group, could have affected the mechanisms by which the UDGs are produced (i.e. tidal interactions, \citealt{2019MNRAS.485..796M, 2019MNRAS.485.1036M}).

The colours of the two groups of UDGs in Antlia also underline their differences. The mean colours are $C-T_1 = 1.58 \pm 0.34$\,mag and $1.81 \pm 0.30$\,mag for G1 and G2, respectively, making  G1 bluer and G2 redder than the CMR fit (see Eq.\,\ref{eq:regression}). Since galaxy colours establish constraints on their formation processes and the Antlia UDGs present colours that are in agreement with the CMR, it is possible that both UDG groups share similar formation mechanisms, but have different evolutive histories due to the subsequent interactions within the cluster potential.

From the theoretical side, \cite{2017MNRAS.466L...1D} studied cosmological simulations of isolated galaxies which include gas outflow processes due to SNe and massive stars; they were able to reproduce the main observed characteristics of UDGs, based on the idea that the internal processes are more efficient than the environmental ones. Their simulations predict colours $B-V = 0.77 \pm 0.12$\,mag, consistent with the mean colour of the G1 population, $B-V = 0.72$\,mag (we used \citealt{2005MNRAS.361..725B} for the $C-T_1$ to $B-V$ colour transformation), and the mean colour of the UDG sample in Fornax \citep{2017A&A...608A.142V}. Additionally, the rest of structural parameters ($r_{e}$, $\mu_{e}$) obtained from the simulations are consistent with the UDGs in Antlia with the exception of the S\'ersic index.

The G1 galaxies, that have the brighter $\mu_\mathrm{e}$ and bluer colours, are distributed on average around $R = 33.35 \pm 9.51$\,arcmin, which is significantly larger than that for G2 ($R = 19.85 \pm 9.07$\,arcmin) that coincides with the distance at which NGC\,3258 is located.

Finally, G1 galaxies have bluer colours than the CMR, are isolated and mainly scattered across the cluster area. These UDGs have similar colours than the simulated galaxies in \cite{2017MNRAS.466L...1D}. On the other hand, the G2 is located in the internal part of the cluster, with mean colours redder than the CMR. One possible explanation is that the natural UDG population of each dominant galaxy have suffered different exposition to the environmental processes, due to the merger between the group centred in NGC\,3258 and the original population.

\section{Conclusions}\label{sec:conclusions}
In this third paper, which is part of a series (\citetalias{2015MNRAS.451..791C, 2018MNRAS.477.1760C}) about the Antlia ETG populations, we have added the surface photometry of $\sim 130$ galaxies in the magnitude range $-21 < M_{V} < -11$\,mag, that have not been previously identified. The majority of the new galaxies occupy the LSB regime; however, some brighter ETGs (mainly S0) confirm the change in slope of the CMR at $M_{T1} \sim -20$\,mag (Fig.\,\ref{fig:CT1_T1}). At the same time, the dispersion of the CMR is not significantly increased with respect to the previous papers. The colour dispersion at each magnitude from $T_{1} = 15$ to $19$\,mag is $\sigma_{(C-T1)} =$ 0.16 - 0.17. The dispersion of the Virgo CMR, in the same luminosity regime, is $\sigma_{(g'-z')} =$ $0.144$\,mag \citep{2019A&A...625A..94H}, which results to be larger than the dispersion obtained for Fornax and Coma. This larger scatter in the Antlia dEs colours and the most bluer galaxies that we found, could indicate that the forming star processes in the LSB population have no finished yet.

We reanalyse the relations between the structural parameters for the Antlia ETGs  and, to complement the low surface brightness regime, compare them with those of the Fornax cluster. 
A linear relation is present between $\mu_{e}$ and $M_{V}$ (Fig.\,\ref{fig:Mv_mue}) that extends from dEs and dSphs up to bright Es and S0, while UDGs and cEs are outliers. The two dominant cD galaxies also fall below the linear fit. 

The projected spatial distribution (Fig.\,\ref{fig:Wong-f1} and \ref{fig:radec1}) shows an elongated distribution of the ETG population in the direction connecting NGC\,3258 and 3268. The smoothed X-ray map \citep{2016ApJ...829...49W} in Fig.\,\ref{fig:Wong-f1} has a perfect match with some galaxies in the central region.
If we consider the whole MOSAIC images including all galaxies of the final sample, i.e. almost 2.6\,deg$^{2}$, the spatial distribution maintains its elongated structure (Fig.\,\ref{fig:radec1}).  

In order to analyse quantitatively the projected distribution, we split the sample into galaxies redder and bluer with respect to the CMR regression, and into dwarf and bright galaxies with respect to $\mu_{0} = 19$\,\mags\ ($V$-band), counting all galaxies along their distance from NGC\,3268, which is adopted as the cluster centre: 

\begin{enumerate}
\item Bluer galaxies are more numerous until the first 200\,kpc (Fig.\,\ref{fig:hdistance_185-1}), which corresponds to the position of NGC\,3258. Further out, the projected densities of bluer and redder galaxies are quite similar. Moreover, the azimuthal distribution (Fig.\,\ref{fig:hdistance_185_pie-1}) reaches a maximum density in the direction towards (and opposite to) NGC\,3258, confirming the elongated structure mentioned above. 

\item The redder galaxies seem to be more concentrated to the bright galaxies in the cluster, and their radial velocities clearly occupy the centre of the histogram between $V_{R} =$ 2000 and 3000\kms.

\item The bright galaxies tend to be concentrated towards the cluster centre, while the azimuthal distribution remains homogeneous. The case is different for the dwarf galaxies, which trace the general trend of the ETG population.
\end{enumerate}

There does not appear to be any colour gradient along the cluster-centric radius (Fig.~\ref{fig:h_185-CT1-re}), being the  mean colour $\langle (C-T_1)_{0} \rangle \sim 1.69 \pm 0.09$\,mag up to 700\,kpc. However, the mean r$_\mathrm{e}$ of the galaxies  decreases from the centre outwards, with a stronger slope outside of the position of NGC\,3258.

The Antlia LF (Fig.\ref{fig:funcion_de_luminosidad}) can be well fitted with a single Schechter function, with a slope comparable with those of the Fornax and Virgo clusters. Additionally, the low number of bright ETGs, the weak X-ray emission in the cluster centre, the ongoing merger of NGC\,3258 group, and the apparently early state of virialization, may be the main causes of the Antlia LF shape and the lack of a dip at intermediate magnitudes.

We found 12 UDG candidates that meet the criteria described by \citet[][]{2015ApJ...798L..45V}. All UDG candidates have $r_\mathrm{e} < 3$\,kpc,  which is not the general case as larger values have been found in  other clusters. UDGs in Antlia can be grouped in two different families, characterised by different mean absolute magnitude and effective surface brightness. These two groups also present different colours and projected spatial distributions, possibly indicating a distinctive origin or evolution mechanism, due to the ongoing merger with the group centred on NGC\,3258. 

\section*{Acknowledgements}
We thank to an anonymous referee for constructive remarks. This work was funded with grants from Consejo Nacional de Investigaciones Cient\'{\i}ficas y T\'ecnicas de la Rep\'ublica Argentina, Agencia Nacional de Promoci\'on Cient\'{\i}fica y Tecnol\'ogica, and Universidad Nacional de La Plata, Argentina. 
LPB, JPCaso and MG: Visiting astronomers, Cerro Tololo Inter-American Observatory, National Optical Astronomy Observatories, which are operated by the Association of Universities for Research in Astronomy, under contract with the National Science Foundation.

\section*{Data availability}
The data underlying this article are available in the article and in its online supplementary material.

\bibliographystyle{mnras}
\bibliography{Antlia_cluster_Paper_3}
\bsp 
\label{lastpage}
\end{document}